\documentclass[showpacs,preprintnumbers,amsmath,amssymb]{revtex4}

\usepackage{graphicx}
\usepackage{dcolumn}
\usepackage{bm}

\newcommand{\numu}{\mbox{$\nu_{\mu}$}\hspace{1mm}}                   
\newcommand{\nue}{\mbox{$\nu_{e}$}\hspace{1mm}}                      
\newcommand{\nutau}{\mbox{$\nu_{\tau}$}\hspace{1mm}}                 

\begin{document}


\title{A first measurement of the interaction cross section of the tau neutrino} 

\author{ 
K. Kodama$^1$, N. Ushida$^1$, 
C. Andreopoulos$^2$, N. Saoulidou$^{2,a}$, G. Tzanakos$^2$,
P. Yager$^3$,
B. Baller$^4$, D. Boehnlein$^4$, W. Freeman$^4$, B. Lundberg$^4$, J. Morfin$^4$, R. Rameika$^4$, 
S.H. Chung$^5$, J.S. Song$^5$, C.S. Yoon$^5$, 
P. Berghaus$^6$, M. Kubantsev$^6$, N.W. Reay$^6$, R. 
Sidwell$^6$, N. Stanton$^6$, S. Yoshida$^6$,
S. Aoki$^7$, T. Hara$^7$, 
J.T. Rhee$^8$,
D. Ciampa$^9$, C. Erickson$^9$, M. Graham$^9$, E. Maher$^{9,b}$, K. Heller$^9$, 
R. Rusack$^9$, R. Schwienhorst$^9$, J. Sielaff$^9$, J. Trammell$^9$, J. Wilcox$^9$,
T. Furukawa$^{10}$, K. Hoshino$^{10}$, H. Jiko$^{10}$,
M. Komatsu$^{10}$, M. Nakamura$^{10}$, T. Nakano$^{10}$, 
K. Niwa$^{10}$, N. Nonaka$^{10}$, K. Okada$^{10}$, B. D. Park$^{10}$, O. Sato$^{10}$, 
S. Takahashi$^{10}$,
V. Paolone$^{11}$, 
C. Rosenfeld$^{12}$, A. Kulik$^{11,12}$, 
T. Kafka$^{13}$, W. Oliver$^{13}$, T. Patzak$^{13,c}$, J. Schneps$^{13}$
\\[\baselineskip]
(The DONuT Collaboration)\\[\baselineskip]
$^1${\it Aichi University of Education, Kariya, Japan}\\ 
$^2${\it University of Athens, Athens 15771, Greece}\\
$^3${\it University of California/Davis, Davis, California 95616, USA}\\
$^4${\it Fermilab, Batavia IL 60510, USA}\\
$^5${\it Gyeongsang University, Chiju, South Korea}\\
$^6${\it Kansas State University, Manhattan, Kansas 66506, USA}\\
$^7${\it Kobe University, Kobe, Japan}\\
$^8${\it Kon-kuk University, Seoul, S. Korea}\\
$^9${\it University of Minnesota, Minneapolis, MN 55455, USA}\\
$^{10}${\it Nagoya University, Nagoya 464-8602, Japan}\\
$^{11}${\it University of Pittsburgh, Pittsburgh, PA 15260, USA}\\
$^{12}${\it University of South Carolina, Columbia,
 South Carolina 29208, USA}\\
$^{13}${\it Tufts University, Medford, MA 02155, USA}\\[\baselineskip] 
$^a${\small Presently at Fermilab}\\
$^b${\small Presently at the Massachusetts College of Liberal Arts,
 North Adams, MA 01247, USA}\\
$^c${\small Presently at the Universit\'e Paris 7, 
Paris, France}\\[\baselineskip]
}


\begin{abstract} 
The DONuT experiment collected data in 1997 and published first results
in 2000 based on four observed $\nu_\tau$ charged-current (CC) interactions.
The final analysis of the data collected in the experiment is presented
in this paper, based on $3.6 \times 10^{17}$ protons on target using the
800 GeV Tevatron beam at Fermilab. The number of observed $\nu_\tau$ CC
interactions is 9, from a total of 578 observed neutrino
interactions. 
We calculated the energy-independent part of the tau-neutrino CC cross section 
($\nu + \bar \nu$),
relative to the well-known $\nu_e$ and $\nu_\mu$ cross sections.
The ratio $\sigma(\nu_\tau)$/$\sigma(\nu_{e,\mu})$ was found to be 
$1.37\pm0.35\pm0.77$. 
The $\nu_\tau$ CC cross section was found to be
 $0.72 \pm 0.24\pm0.36 \times 10^{-38}$ cm$^{2}\rm{GeV}^{-1}$.
Both results are in agreement  with expectations from the Standard Model.
\end{abstract} 


\maketitle 

\centerline{\today}

\large 

\section{Introduction} 

The tau neutrino, $\nu_\tau$, was assigned its place in the Standard Model
after its electrically charged weak isospin-$\frac{1}{2}$ partner, the $\tau$ lepton,
was discovered in 1975 \cite{ref:perl}.
The observation of identifiable $\nu_\tau$ interactions,
in a manner similar to $\nu_e$  \cite{ref:reines} and
$\nu_\mu$ \cite{ref:lederman} interactions, did not immediately follow.
The difficulty of measuring $\nu_\tau$ interactions
was due to the relative scarcity of the sources of $\nu_\tau$ and
the lack of sufficiently powerful detection methods to unambiguously identify
the short-lived $\tau$ lepton (mean lifetime  $2.9 \times 10^{-13}$ s)
 produced in $\nu_\tau$ charged-current interactions.
 These challenges were overcome in the observation of four $\nu_\tau$
 interactions by the DONuT ({\bf D}irect {\bf O}bservation
of {\bf Nu}-{\bf T}au) collaboration, in 2000 \cite{ref:DONuTfirst}\cite{ref:AnnRev},
 twenty-five years after the $\tau$ was discovered.
 Analysis of our full data set yielded
nearly three times as many neutrino interactions of all flavors as reported in 
Ref.  \cite{ref:DONuTfirst}. This paper reports our final results, bringing the
DONuT experiment to a completion.

The purpose of the DONuT experiment was to study
$\nu_\tau$ charged-current (CC) events,
\begin{subequations}
\label{eq:tauCC}
\begin{align}
\nu_\tau + N  \rightarrow  \tau^- + X, \\
\bar\nu_\tau + N  \rightarrow  \tau^+ + X.
\end{align}
\end{subequations}
 However, during data taking, DONuT was recording interactions of neutrinos of all flavors: 
\noindent $\nu_e$ CC events
\begin{equation}
   \nu_e + N \rightarrow e^- + X,
\label{eq:eCC}
\end{equation}
$\nu_\mu$ CC events
\begin{equation}
   \nu_\mu + N \rightarrow \mu^- + X,
\label{eq:muCC}
\end{equation}
and neutral-current (NC) events
\begin{equation}
 \nu_\ell + N \rightarrow \nu_\ell + X,\ \ \   \ell=e,\mu,\tau
\label{eq:NC}
\end{equation}
and analogously for the antineutrinos.

Reaction (\ref{eq:tauCC}) must be distinguished from charm production
in reactions (\ref{eq:eCC}) and (\ref{eq:muCC}), since the tau-lepton and
the charmed particles have comparable lifetimes and decay signatures:
\begin{equation}
   \nu_\ell + N \rightarrow \ell^- + C^\pm + X,\ \ \   \ell=e,\mu
\label{eq:charm}
\end{equation}
where  $C$ = D, D$_s$, or $\Lambda_c$.
Another background considered here were secondary hadron ineractions
in NC neutrino events, reaction (\ref{eq:NC}),
\begin{eqnarray}
\label{eq:scatter}
\nu_\ell + N \rightarrow\ \nu_\ell+ h^\pm + X,\ \ \    \ell=e, \mu, \tau, \\
{\rm followed\ by}\ \ \ h^\pm + N \rightarrow ({\rm 1\ or\ 3\ prongs}) + X^0 \nonumber
\end{eqnarray}

The experimental apparatus and techniques, have been described in detail
elsewhere \cite{NIM-emul}\cite{NIM-spect}and are only summarized here.

The location of vertices in the emulsion data, tagging leptons and the
subsequent search for secondary vertices, were accomplished with high
efficiency. This allowed a detailed event-by-event analysis with small
and calculable background levels. Further, the large amount of information
in the emulsion/spectrometer system permitted the use of powerful
multivariate methods yielding probabilities  for each candidate event to be
signal or background. The measured $\nu_\tau$ cross section was computed
using the final sample of all $\nu_\tau$, $\nu_e$, and $\nu_\mu$
interactions located in the emulsion.

The organization of this paper is as follows.
First we give an overview of the neutrino beam and detector elements.
Next, there is a synopsis of triggering and filtering that produced the interaction sample.
We then give important details of the emulsion detector.
The analysis is reviewed by outlining the lepton identification procedures, the
Monte Carlo, event location in the emulsion and secondary vertex search.
After a survey of the entire data set including neutrino interactions of all flavors,
the $\nu_\tau$ cross section analysis is described, systematic error sources are discussed,
and the results are presented.

\section{Neutrino beam and detector}
\label{sec:beamdet}

{\it Primary beam.} The number of 800-GeV protons that struck the beamdump
 was measured by devices that integrate charge
 collected from secondary emission from a foil.
 These monitors were calibrated with a beta source
 before the experiment began. Several times during
 the course of the run, these devices were
 calibrated against coil pickups and other
 monitors installed in the accelerator extraction
 complex. These checks showed that the primary
 beam monitors were consistent within 5\% at
 intensities of $5 \times 10^{12}$ to $1 \times 10^{13}$ protons per
 spill. Losses in the beamline were small
 ($\approx 10^{-5}$), and no other corrections were applied.
The monitors' output was digitized and recorded at the
 experiment, and gated by the trigger electronics.
 A total of $3.54 \times 10^{17}$ protons were recorded during
 the live-time of the experiment. A systematic
 uncertainty of 5\% was assigned to the value of
 the total number of protons in the beamdump.

\begin{figure}
\includegraphics[width=\columnwidth]{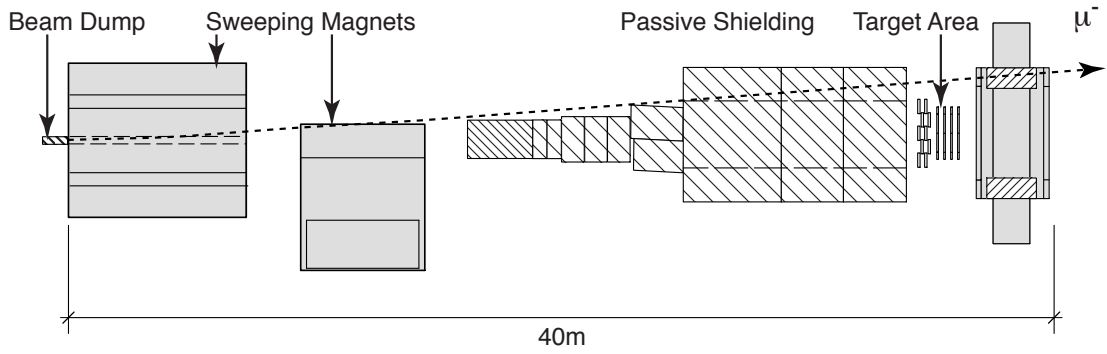}
\caption{
Schematic plan view of the neutrino beam. The 800 GeV protons are incident on the beamdump from the left. The emulsion modules are located within the target area, 36 m from the beamdump. The trajectory of a 400 GeV/$c$ negative muon is shown. Note that the passive steel shield does not fill the volume occupied by high-energy muons along the plane of the beamline.}
\label{fig:beam}
\end{figure}

{\it DONuT beamline.}
The 800-GeV protons from the Tevatron were stopped in a beamdump
in the form of a solid block of tungsten alloy. The typical
intensity was $8 \times 10^{12}$ protons for 20 seconds each minute,
or about 20 kW of  beam power. Immediately following the
beamdump were two dipole magnets with solid steel poles, providing both
absorption of interaction products and deflection of high-energy muons
away from the beam center. Following the magnets was an additional
18 m of passive steel shielding limited to within 2 m of the beamline.
Emerging at the end of this shield, 36 m from the beamdump,
 were neutrinos and muons. The muons
were mostly contained in horizontal fan-like distributions on each side
of the centerline. The neutrino beam design is shown in Fig. \ref{fig:beam}.

\begin{figure}
\includegraphics[width=\columnwidth]{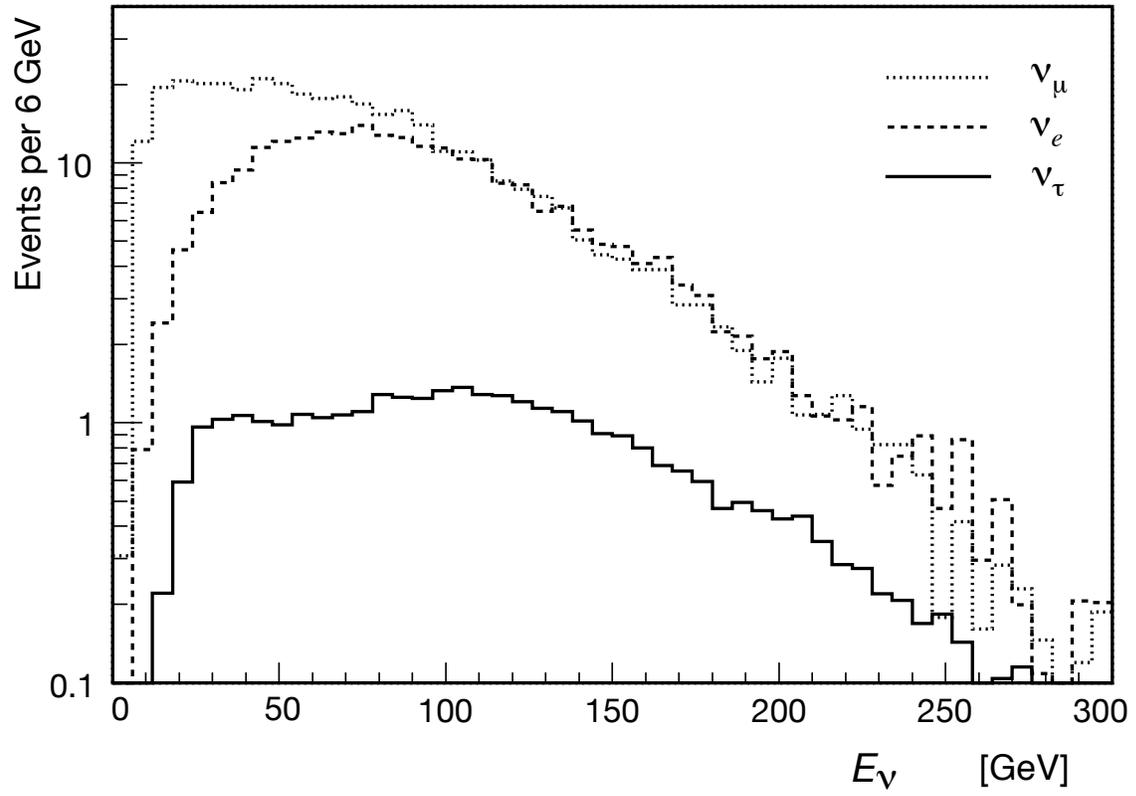}
\caption{
Calculated energy spectra of neutrinos interacting in the DONuT emulsion target. }
\label{fig:nu_spectrum}
\end{figure}

{\it Neutrino beam.}
Neutrinos in the DONuT beam originated from decays of particles within
the hadron shower created by a primary proton interaction. Neutrinos
from decays of charmed particles are called {\it prompt} neutrinos,
and neutrinos from decays of  $\pi^\pm$ and K$^\pm$ are called
{\it non-prompt} neutrinos. About 97 \% of the neutrino flux from the
beamdump was composed  of $\nu_e$ and $\nu_\mu$, the rest being $\nu_\tau$.
 93\% of the $\nu_e$'s were prompt, while $\nu_\mu$'s had substantial
components of both prompt and non-prompt neutrinos.
All $\nu_\tau$'s were prompt. Most of them originated in leptonic decays
of D$_s$ mesons. The decay mode D$_s \rightarrow \nu_\tau \tau$ 
yielded two tau neutrinos within a distance of a few millimeters.
This decay length is much less than the interaction length of six centimeters.
 The calculated neutrino energy spectra of all the
 neutrinos that interacted in the DONuT target are shown in
 Fig.\ \ref{fig:nu_spectrum}.

\begin{figure}
\includegraphics[width=\columnwidth]{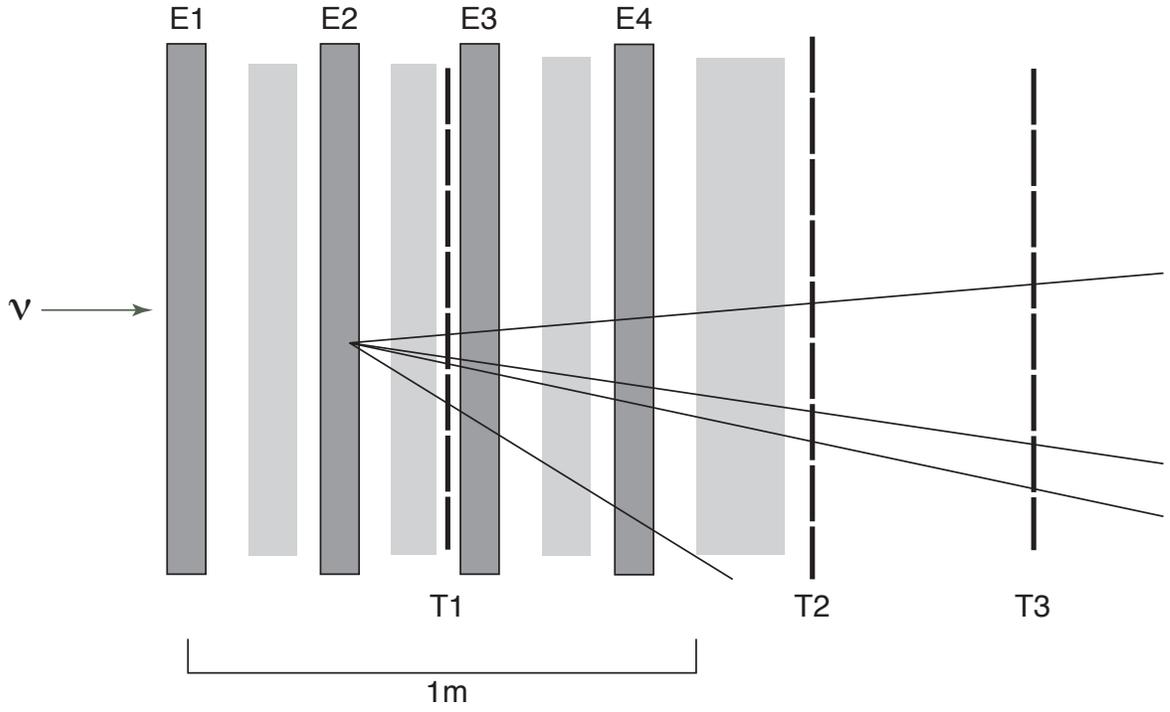}
\caption{
Schematic plan view of the target region. The emulsion modules are indicated with ïEÍ labels, the trigger
hodoscopes with ïTÍ labels. The lighter gray areas are occupied by scintillating fiber planes, 44 in total. The paths of charged particles in a typical interaction are superimposed.}
\label{fig:SFplanes}
\end{figure}

{\it Emulsion target.} The target -  schematically depicted in
 Fig. \ref{fig:SFplanes} - was the
core of DONuT. Its capabilities and performance were matched to the
task of recognizing neutrino interactions containing tau leptons.
The main component of the target assembly was 250 kg of nuclear emulsion
stacked in modular fashion along the beamline.
A total of seven emulsion modules in the target station were exposed,
with a maximum of four modules in place at any time during the experiment.

Each module was exposed for a limited time to avoid track density higher than
$10^5$ tracks per cm$^2$ that would make the emulsion data analysis
 inefficient. To further assist the analysis, single
{\it Changeable Sheets} were mounted 1 cm downstream of each emulsion target
module and replaced ten times more often.

{\it Scintillating fiber tracker (SFT)}.
Integrated into the emulsion target station were 44 planes of the SFT
built using 0.5-mm-diameter scintillating fibers to provide medium-resolution
tracking and a time-stamp for each event.

\begin{figure}
\includegraphics[width=\columnwidth]{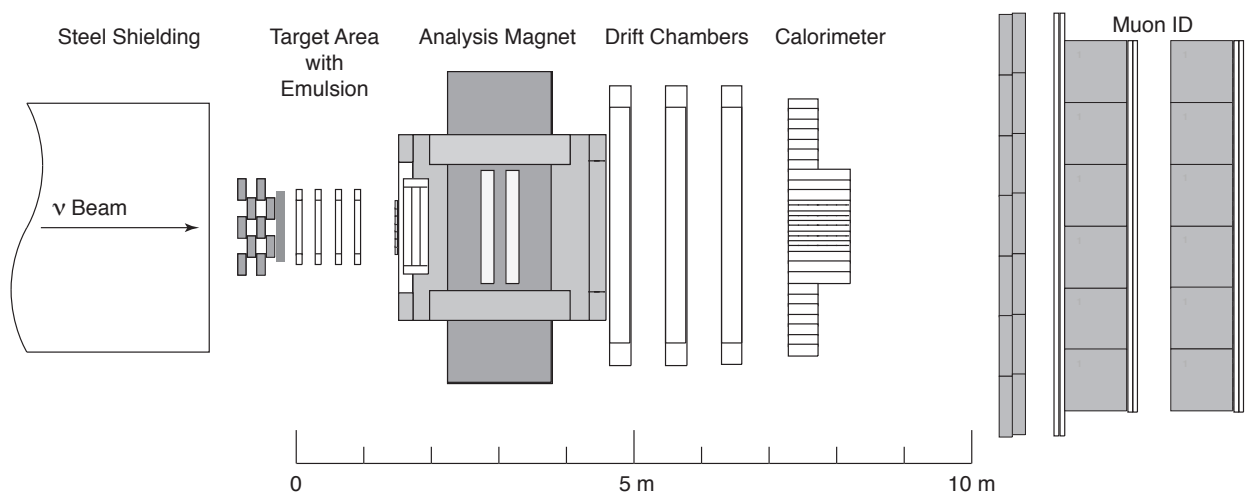}
\caption{
Schematic plan view of the spectrometer. The neutrinos are incident from the left, emerging from the passive shield. The design is relatively compact, to optimize identification of leptons (muons and electrons).}
\label{fig:spect}
\end{figure}

{\it Spectrometer.} The emulsion target station was followed by a spectrometer
consisting of a large-aperture dipole magnet and up to six
{\it drift chambers}. A lead- and scintillating-glass
{\it electromagnetic calorimeter} aided in identifying electrons and
measuring their energy. Behind the calorimeter, muons were tagged with a
{\it Muon-ID} system consisting of three steel walls each followed by
two crossed proportional-tube planes.
The plan of the spectrometer is shown in Fig. \ref{fig:spect}.

\section{Spectrometer data collection and reduction}
\label{sec:dataredu}

\subsection{Triggering and data acquisition}
\label{sec:trig}

{\it Trigger.} A trigger for recording neutrino
interactions required that no charged particles entered the emulsion from
upstream and at least one charged particle emerged from an emulsion target.
The scintillation-counter triggering system included
a veto wall upstream of the emulsion target and three hodoscope
planes distributed between and downstream of the emulsion modules, shown in
Fig.\ \ref{fig:SFplanes}.
The average trigger rate was 5.0 Hz, with a livetime of 0.89.
The trigger efficiency was calculated using simulated neutrino interactions
and measured efficiencies for all counters. 
The efficiency for triggering on $\nu_e$ CC, $\nu_\mu$ CC, $\nu_\tau$ CC,
 and NC interactions was 0.98, 0.96, 0.96, and 0.86, respectively.
 Detailed description of the triggering system can be found
in Ref. \cite{NIM-spect}.

{\it Data acquisition.} The architecture of the data aquisition was based
 on the Fermilab DART product \cite{ref:dart},
 using VME-based microprocessors to control
 the transport of data from the VME buffers to a host computer.
 The host computer served as both the data monitor and
 as the data logger to tape (Exabyte 3500). The average event size
 was 100 kB, with a throughput of 10 MB per beam cycle of one minute.

\subsection{Filtering and scanning}
\label{sect:filt}

A total of $6.6 \times 10^6$ triggers from $3.54 \times 10^{17}$
 protons on target were recorded.
 In this data set, only about $10^3$ neutrino interactions
were expected. This implied that the great majority
of the triggers were background processes satisfying the simple trigger
requirements of Section \ref{sec:trig}.
 Data from the electronic detectors were used
to extract the neutrino interaction candidates in a two-step process.

{\it Software filter.}
The time difference between any two trigger counter
signals was required to be within 2.5 ns.
Data from the SFT and from the drift chambers were then used to
reconstruct tracks and to search for a vertex near one of the emulsion
targets. Triggers that did not yield a candidate vertex were eliminated.
This software filter reduced the number of recorded triggers by
 a factor of 300.
 Efficiencies for keeping neutrino interactions were determined by
Monte Carlo studies to be 0.98 (for CC events) and 0.96 (for NC events).

{\it Physicist scan.}
In the second step, the remaining triggers were scanned individually by a
physicist using a graphical display. This step rejected events
originating from particle showers produced by high-energy muons and
checked for errors in reconstruction and other pathologies. Most
 of the events were rejected quickly and with high confidence.
This visual scanning reduced the data by another factor of 20,
yielding 866 neutrino interaction candidates within one of the
emulsion modules which had a visible energy over 2 GeV.
The efficiency of the physicist scan was found to be $(0.86 \pm 0.07)$.

The estimated total efficiency for retaining a $\nu_\tau$ CC
interaction with the electronic detectors was 0.72 after
triggering, filtering and scanning. For $\nu_e$ ($\nu_\mu$) CC interactions
these efficiencies were 0.73 (0.71), and for NC interactions it was 0.64.

\subsection{Neutrino event sample}
\label{sec:evtsampl} 

 The resulting sample included 866 events that were  likely
neutrino interactions of all flavors with the
vertex located within the fiducial volume in the emulsion target.

We report here on the analysis of all the events for which the neutrino
interaction vertex was found in the emulsion, referred to thoughout
 as {\it located events}.
Although locating the vertex in the emulsion was attempted for each of the
866 events, only 578 events were located, as described in
 Section \ref{sec:evt-loc}.

Events in the initial sample that were not located in the emulsion
were not used in the analysis described below.

\section{The emulsion}

\label{sec:emulsion}

\begin{figure}
\includegraphics[width=\columnwidth]{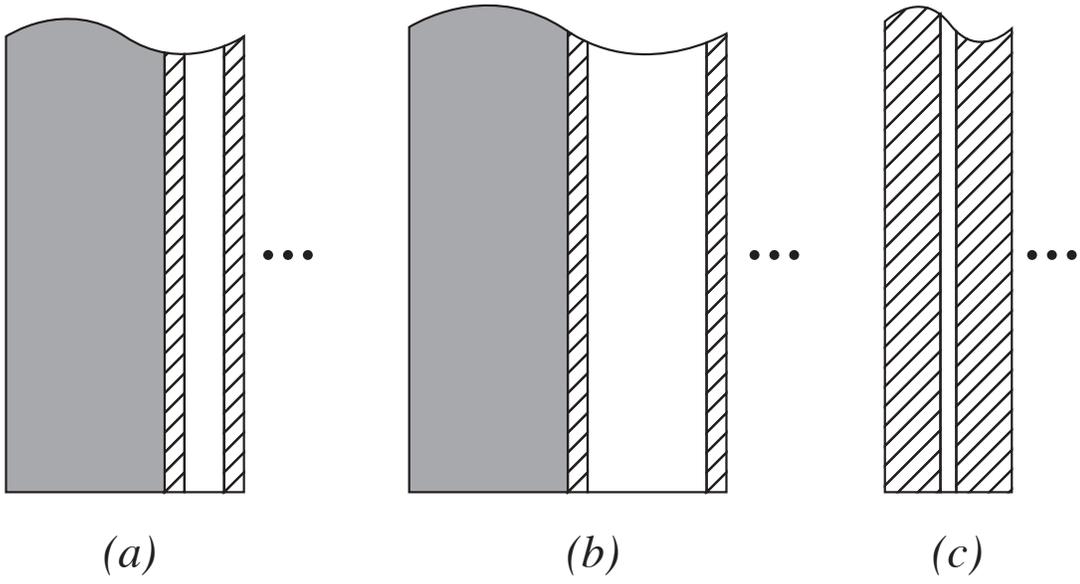}
\caption{
Emulsion target designs. The ECC designs (a) and (b) used 1-mm thick stainless steel sheets interleaved with emulsion plates using 100 $\mu$m thick emulsion layers on a 200-$\mu$m plastic base in (a), and 800-$\mu$m plastic base in (b).  Most neutrino interactions were in the steel. The bulk emulsion type (c) used 350-$\mu$m emulsion layers on 90-$\mu$m plastic base, without steel. Steel is indicated by shading, emulsion by cross-hatching, the plastic base is unshaded.}
\label{fig:emul}
\end{figure}

The DONuT emulsion modules were the first modern implementation of a
design that interleaves metallic sheets (stainless steel) with emulsion
sheets to achieve high mass to increase the number interactions and
 high precision for tau recognition. As illustrated in Fig. \ref{fig:emul},
two designs of these `Emulsion Cloud Chambers' were used in DONuT:
both used 1-mm thick steel sheets interleaved with emulsion sheets
 having 100 $\mu$m thick emulsion layers on both sides of a plastic base.
The designs differed in thickness of the base, one was 200 $\mu$m and
the other 800 $\mu$m thick. The third design had 350 $\mu$m thick emulsion
layers on 90 $\mu$m thick base. More details about the emulsion target design
can be found in Ref. \cite{NIM-emul}.

After exposure, the emulsion target modules were transported to
Nagoya University in Japan, where they were disassembled
and individual emulsion sheets developed.
The Changeable Sheets were developed at Fermilab.

The information from a small emulsion volume surrounding the
interaction point predicted by the spectrometer data was fully digitized and
used  in a manner similar to the information from an electronic
detector. The size of the volume needed to be large enough
to contain the vertex but small enough to be compatible with
the capabilities of the emulsion scanning machines.

Once the desired emulsion volume was determined, the individual
emulsion sheets were digitized using automatic scanning and
digitizing apparatus at Nagoya University. The Nagoya group developed
this technology over the years, starting in 1974.
The DONuT emulsion data were obtained using
Ultra Track Selector (UTS) digitizers \cite{NIM-Aoki} with scanning rate
of 1 cm$^2$/hour, a factor of five  improvement over the
technology used to obtain the first DONuT results of
Ref. \cite{ref:DONuTfirst} allowing for greatly increased location efficiency.

{\it Emulsion data.}
 The UTS automated scanning stations found and digitized track segments 
(``microtracks") in the emulsion layers on both sides of the
 transparent plastic base. Both the position and angle of each segment
 were computed and recorded in real time. Efficiency for detecting
microtracks was measured to be greater than 0.97.

Complete tracks were built layer by layer. Each microtrack was examined
 to see if it had a connectable microtrack
in adjacent emulsion layers. Once reconstructed, the tracks were added
to a data set unique to the given scan volume.

An important tool used in the offline emulsion data processing were
high-energy muons from the beamdump that penetrated
the shielding and were recorded in the scanned emulsion volume
 as through-going tracks with little measurable scattering, called
``calibration tracks" below.

{\it Data quality checks.} A systematic methodology was
developed to quantify the quality of tracks found in digitized emulsion images.
Two quantities were used:
(a) position accuracy $\sigma$ as measured by rms displacement of microtracks
 from fitted calibration tracks, and
 (b) emulsion read-out efficiency $\varepsilon$, representing the fraction of
 identified calibration-track microtracks actually seen in
 any one emulsion plate.
Emulsion data passed the data quality check when
$\sigma \le 1.0\,\mu$m, and $\varepsilon \ge 0.9$.
 Reasons for poor data quality could be a damaged emulsion (lost forever),
 difficulty in emulsion digitization (to be re-digitized), or
 a systematic problem such as emulsion-sheet slipping
within a stack which can be corrected as detailed below.
More than 50\% of events where the predicted vertex was not initially
found in the emulsion fell into the poor-data-quality category.

{\it Emulsion-sheet slipping:}
 Occasionally, emulsion sheets slipped one with respect to another
during exposure. An alignment method was therefore devised to correct
 for it using the calibration tracks.
 The alignment parameters of interest included the
 distance between the emulsion layers, the relative shifts in
 transverse direction and the shrinkage of the emulsion layers.
 Alignment between adjacent sheets was determined within 0.2 $\mu$m.

\section{Particle identification}

\subsection{Muons}

\begin{figure}
\includegraphics[width=\columnwidth]{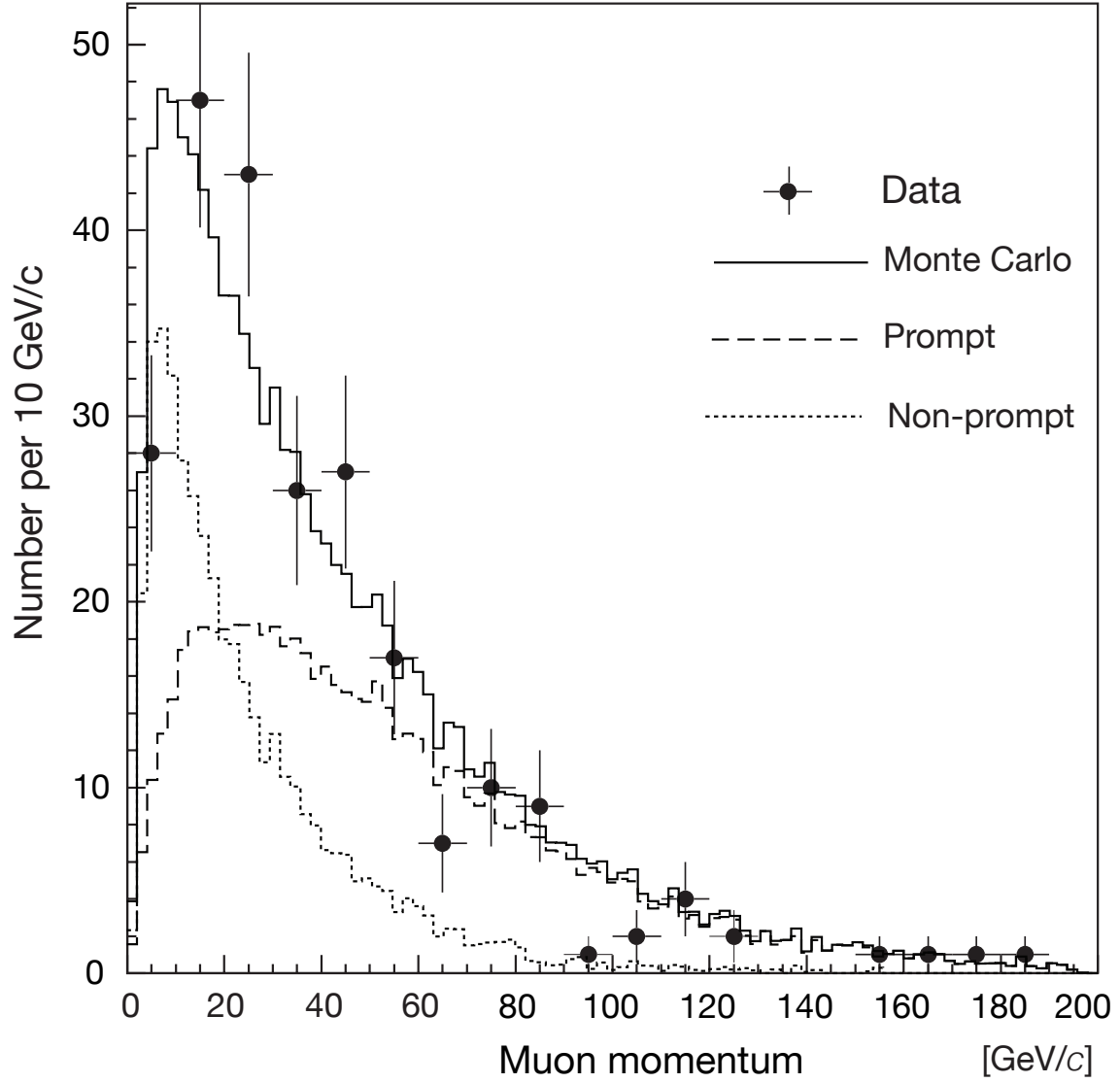}
\caption{
Muon momentum distribution for  the set of 578 located events.
The data are shown by solid circles, and Monte Carlo expectation is the solid histogram.
Also shown are the expected muon distributions from the two components of the $\nu_\mu$ flux,
prompt (dashed) and non-prompt (dotted) histograms.}
\label{fig:muspec}
\end{figure}

A muon tag was assigned to a track if there were at least four hits
in the six proportional-tube planes of the muon-ID system.
 The per-tube efficiency for muons was measured
 to be 0.96, and the geometrical acceptance of the muon ID
 system was estimated by Monte Carlo to be 0.76, yielding
 an overall efficiency of 0.73. The muon spectra are shown 
 in Fig.\ \ref{fig:muspec}.

Muon track momentum could be measured in one of two very different ways:
{\it (i)} from the curvature in the spectrometer, and
 {\it (ii)} from multiple coulomb scattering (MCS) in the emulsion.

{\it Spectrometer measurement.} In the spectrometer, track momentum
was measured using a 4 T  magnet with $\int B dl$ = 0.75 T m.
For muons, $\Delta p/p$ was 11\% for momentum $p$ of 20 GeV/$c$,
increasing to 100\% at $p$ = 250 GeV/$c$.

{\it Emulsion measurement.}
The high spatial precision of the tracking in emulsion, in conjunction with
an adequate sampling rate,  allowed the calculation of  track momentum from
 the visible scattering of the track's segments (microtracks)
 in individual emulsion plates.

A special emulsion track scan was performed on all tracks found
in candidate neutrino events for the dual purpose of the multiple coulomb
scattering measurement and
 electron identification (see Section \ref{sec:eID} below).
Momentum was successfully measured using multiple coulomb scattering
 for 64\% of the tracks in the sample.

\begin{figure}
\includegraphics[width=\columnwidth]{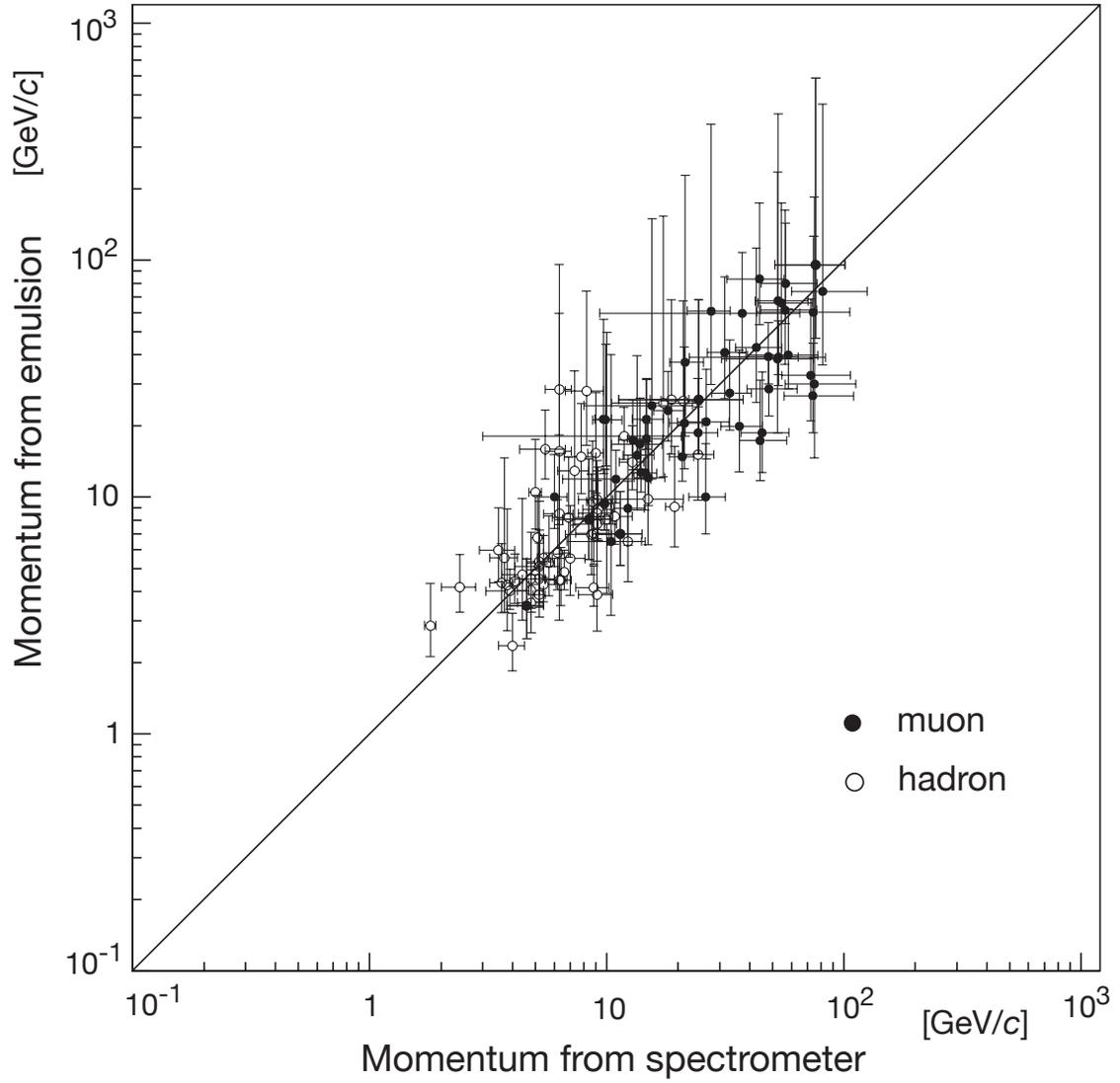}
\caption{
A comparison between track momenta measured by multiple coulomb scattering in emulsion
and by the spectrometer. Although the tracks tagged as muons avoid secondary
interactions, the momenta are often at upper limit of measurement in the emulsion.}
\label{fig:mcs_spect}
\end{figure}

The method was validated by test-beam experiments which showed that the beam
momentum of 0.8 and 1.5 GeV/$c$ (4 Gev/$c$) could be measured by the emulsion
 with a resolution of 23\% (30\%) \cite{NIM-Park} (\cite{NIM-emul}).
 A comparison of
 track momenta measured with both the emulsion and spectrometer
 is shown in Fig.\ \ref{fig:mcs_spect}.

The upper limit of the momentum measured  this way was
 determined by the number of samples, the angle of the track, the
 quality of the emulsion data and the type of emulsion module.
 A typical upper limit was 25 GeV/$c$.


\subsection{Electrons}

\subsubsection{Electron identification}
\label{sec:eID}

The electron analysis was less straightforward since it involved
 several systems. Since the emulsion modules were two to three
 radiation lengths thick, most events
containing electrons would exhibit showers in the SFT and in the
electromagnetic calorimeter. These two electronic detectors were
used to find the most likely initial energy of the electron from an algorithm
using both energy (pulse height) and geometrical shower development.

A special {\it electron ID scan} was performed on all emulsion tracks.
This scan followed each track from the vertex to the most downstream plate.
An area of 600 $\mu$m $\times$ 600 $\mu$m centered on the track
was digitized in each emulsion plate. Electrons were identified by
 electron-positron pairs found within 20 $\mu$m of the track.
 The electron-ID scan was most effective
 for vertices located in the upstream part of an emulsion
 module.

 The efficiency for electron tagging using the spectrometer was estimated
 to be $0.80\pm0.04$.
 The electron tagging efficiency using emulsion data varied
 with path length, with a maximum of  0.86 for tracks passing through
 at least 2 $X_0$.  The  integrated efficiency of identifying an electron in 
the emulsion was 0.66.

\begin{figure}
\includegraphics[width=\columnwidth]{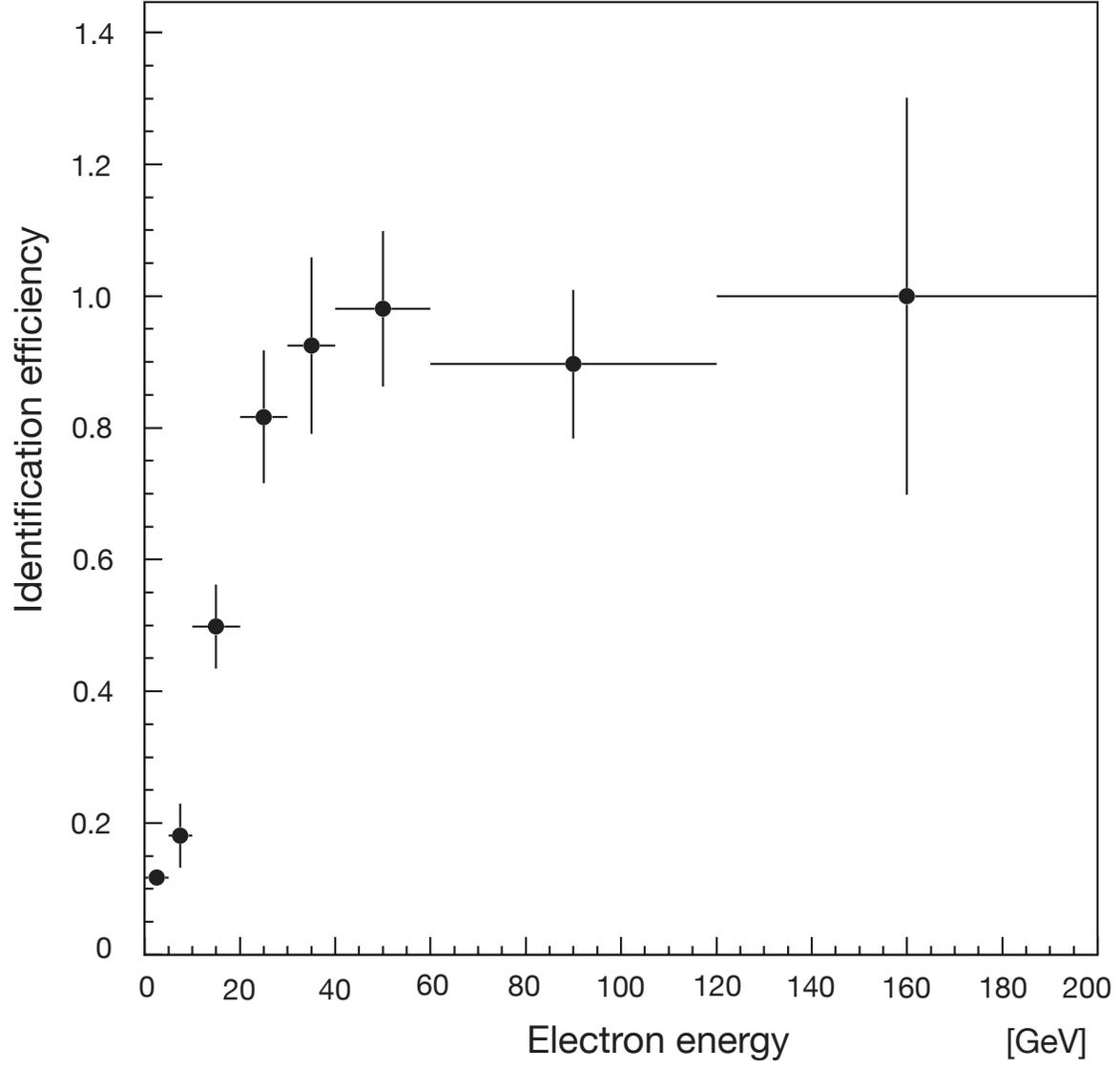}
\caption{
The electron identification efficiency as a function of electron energy. This analysis used the scintillating
fiber detector and the calorimeter.}
\label{fig:eid_eff}
\end{figure}

The total electron identification efficiency as a function of energy
 is shown in Fig. \ref{fig:eid_eff}.
  
\subsubsection{Electron energy measurement}
\label{sec:e_energy}

The target/fiber system was
 also used to estimate the electron (or gamma) energy.
 Since the scintillating fiber system response was
 calibrated to minimum ionizing particles, the total
 pulse height in a shower could be summed for each
 station providing a direct measure of energy. The
 energy estimates at each station were input variables
 for an algorithm to compute electron energy from
 shower development. The calorimeter information
 was added for showers that penetrated less than
 six radiation lengths of emulsion (approximate
 shower maximum). The estimated energy resolution, $\Delta E/E$,
  was 30\%. 

Since the beamline could not be configured for transport of electrons,
 electron identification and energy estimate relied heavily
 on Monte Carlo simulation. A selection of probable
 electrons from interactions in the most downstream emulsion-target module,
 analyzed for momentum in the spectrometer and
 energy in the calorimeter, showed that the
 calorimeter calibration was consistent with
 a calibration method using muons as minimum ionizing particles.

\section{Monte Carlo simulation}
\label{sec:MC}

The production of neutrinos in the beamdump, their transport through the
shielding system, and their interactions in the emulsion target were
simulated with a GEANT3-based Monte Carlo software.
The emulsion target and all electronic detectors in the spectrometer
were simulated, taking into account their measured efficiencies and
other  response characteristics
 peculiar to each system.

The production of charmed particles by 800 GeV protons in the beamdump
were generated using a phenomenological formula,
\begin{equation}
\frac{{d^2 \sigma }}{{dx_F dp_T^2 }} = A\,e^{ - bp_T^2 } \left( {1 -  |x_F| } \right)^n
\label{eq:charm_cs}
\end{equation}
where $x_F$ is Feynman $x$ and $p_T$ is transverse momentum. The  values of
$b$ and $n$ in Eq. \eqref{eq:charm_cs} depend on the charm species. 
The details of the simulation of neutrino production in the beamdump via charm
particle decays are given in Appendix \ref{sec:Charm_production}.
If the path of a neutrino originating from a charm decay intersected the
emulsion target, a deep-inelastic neutrino-nucleon interaction was
generated using LEPTO v6.3.

The simulated particles from the interaction were recorded in each
detector and ``digitized" as appropriate for electronics used in the
experiment. This Monte Carlo data was stored in the format used by
the data acquisition system and was analyzed in the same manner as
experimental data. In addition, a separate file was generated with
data from the charged particles within the emulsion sheets. The data
contains microtracks in each emulsion layer, but it  does not directly
simulate the algorithms used in the UTS emulsion digitizers.

The Monte Carlo was the primary tool for computing acceptance of the
neutrino flux in the emulsion target needed for the cross section analysis.
It was also used to establish selection cuts, develop electron identification
algorithms, and probe systematic effects from charm particle production
uncertainties.

\section{Event location in the emulsion}

\label{sec:evt-loc}

 Two methods were used by DONuT to locate
neutrino inetraction vertices in the emulsion target, both starting with
extrapolation of spectrometer tracks back to the emulsion target.
The SFT was the principal device for making the initial vertex prediction.

\subsection{Event location by Netscan}
\label{sec:Netscan}

Netscan event location was a multi-step process.
Initially, information from the electronic detectors was used to fit
 charged-particle tracks, and reconstruct a neutrino-interaction
 vertex whenever possible.
 The resolution of these detectors enabled vertex predictions
 with a precision of about 1 mm transverse and 5 mm along the
 neutrino beam direction.
Next, both the position and size of the scanning
 volume were determined using the spectrometer prediction,
and all microtracks within the scanning volume were digitized.

After the necessary alignment of the emulsion data,
track pairs were examined to see if they formed a vertex.
The following selection criteria were applied:
\begin{itemize}
\item {Tracks must start within the volume and
cannot be connected to any aligned microtracks in two adjacent upstream
 emulsion layers to reject penetrating muon tracks.}
\item {Tracks must be constructed from at least three microtracks and
 have a good $\chi^2$ fit. These requirements reduce the number of
 low momentum tracks.}
\item {The remaining tracks were tested for vertex topology. Tracks 
  were associated when the impact
 parameter at the best vertex position was less than  5 $\mu$m.}
\end{itemize}
Out of the total of $\sim 10^4 - 10^5$ microtracks per
$ 5 \times 5 \times 15$ $\rm{mm}^3$ emulsion volume, only a few vertex candidates
 remained after the three requirements were imposed.
 To confirm a vertex candidate, (\emph{i}) the emulsion plates
 near the vertex point were examined by a physicist using a manually controlled
 microscope to check for consistency of the neutrino interaction
 hypothesis (i.e. neutral particle interaction), and (\emph{ii}) the emulsion
 track information was compared with the hits in the SFT to verify that
 all tracks were associated with the same event. For interaction vertices that
passed all the checks, all tracks in the event were refit using the emulsion
information.

 \subsection{Event location by Backscan using Changeable Sheets}
\label{sec:CS}

 The Changeable Sheets were used when the vertex prediction was problematic:
 the event was either too complex to have an accurate vertex prediction made,
 or, on the other hand, only one charged track was
 reconstructed in the SFT, so that the
 interaction point was constrained only in the two transverse dimensions.
 In this case, the SFT track was extrapolated to the CS position and
 the emulsion data in this sheet was searched for a track matching
 both position and angle. If found, the track could be followed 
 into the emulsion target module with much greater accuracy to greatly
 reduce ambiguity in high track-density regions. The SFT-CS matched tracks were
 followed upstream, through the sheets of the target module,
 using emulsion scanning within a cylindrical volume (used in Ref.
\cite{ref:DONuTfirst}) or within a conical volume with transverse
 dimensions increasing along the track, used in this analysis.
 The latter scan resulted in much larger emulsion volume
 being scanned to increase event location efficiency, but also
 greatly increased the digitizer work load. This was only possible
 when UTS digitizers became available.
 
If a track penetrated all the way to the most upstream sheet, the track was
 rejected. If the track was found to be missing in upstream sheets,
 it was assumed to originate in the space between emulsion layers.
 All tracks followed in this way were checked to ensure that they
 did not originate as an e$^+$e$^-$ pair, a secondary interaction or 
 as an emulsion inefficiency causing a gap in a throughgoing track.
 If these background hypotheses were rejected,
 the track was assumed to originate from a primary vertex of a neutrino
 interaction. All other emulsion tracks that passed within 5 $\mu$m of
 this track's endpoint were checked to see if they were likely
 to originate in the same interaction.

\subsection {Special cases}

Special methods were developed for events with large number of hits
in SFT, for which the total pulse height exceeded the equivalent of 
650 minium ionizing tracks and no 3-D tracks could be reconstructed.
These large-pulse height events are called {\it LP events} below.

In the {\it modified CS scan}, a large area ($>$ 1 cm$^2$) was scanned
in the CS nearest to the upstream end of a large SFT shower, and
electron signature was searched for in the form of clustered
parallel microtracks. If found, the electron was followed by backscan
to the vertex. Alternatively, a line was drawn through the shower core
in the SFT to better pinpoint the CS area to be scanned, with a typical size
of 5$\times$5 $\rm{mm}^2$. In this case, no electron signature was required, and all
tracks matching the line in position and angle were followed back.

In the {\it modified Netscan}, a number of lines were drawn in $u$- and $v$-projection
and extrapolated into the emulsion module. If  a candidate vertex region
was found, Netscan was applied over an oversized volume,  typically
$13 \times 13 \times 20 $ $\rm{mm}^3$.

The two methods yielded similar numbers of events, with 
a total of 58 LP events located in the emulsion, of which 31 were $\nu_e$ events,
9 $\nu_\mu$ events, 2 $\nu_\tau$ events and 16 NC events.

\subsection{Location efficiency}
\label{sec:loceff}

The overall efficiency for locating the primary vertex in the
 emulsion was given directly as the ratio of the number
 events found and the number of events tried. This ratio
 is 578/866 or $0.667 \pm 0.036$.

We note that each module corresponded to 2.5 to 3 radiation lengths and
0.2 interaction lengths, so secondary interactions
were a common occurrence. Resulting large hadron/electromagnetic showers
hampered track reconstruction and vertex location. 
There were 188 events classified as LP events, or 22\% of the total of 866.
A total of 58 LP events were located in the emulsion, representing a
location efficiency of $0.31 \pm 0.05$, to be compared to $0.77 \pm 0.04$ 
location efficiency for the regular events (520 located out of a total of 678).

\begin{figure}
\includegraphics[width=\columnwidth]{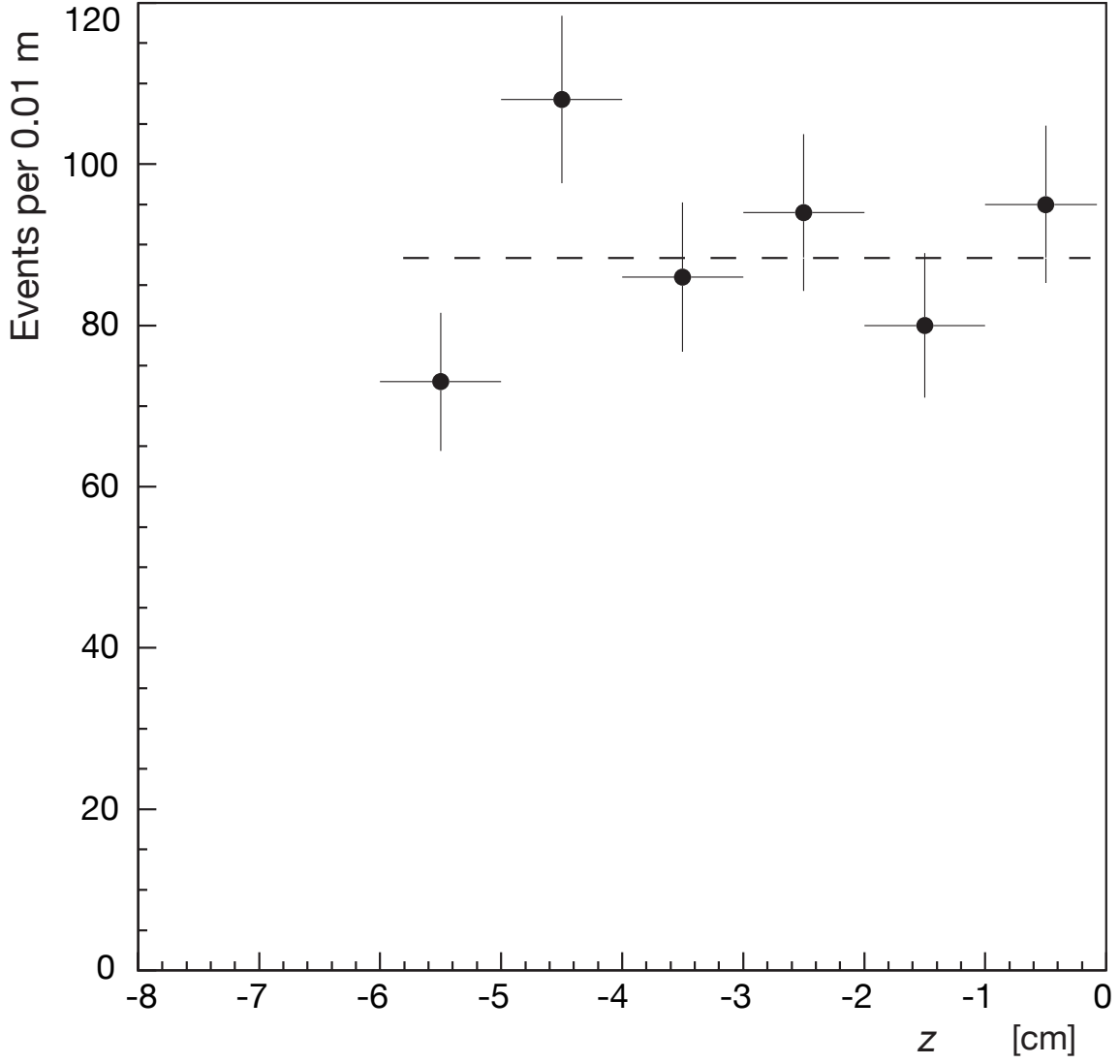}
\caption{
Number of located events as a function of $z$, the vertex position measured from the downstream edge of a module along the beam direction. Data from all seven modules are included. 
Also shown ({\it dashed line}) is
the fit assuming the results are independent of $z$, yielding a value of 88 with $\chi^2/dof$ equal to 1.7.}
\label{fig:z}
\end{figure}

\begin{figure}
\includegraphics[width=\columnwidth]{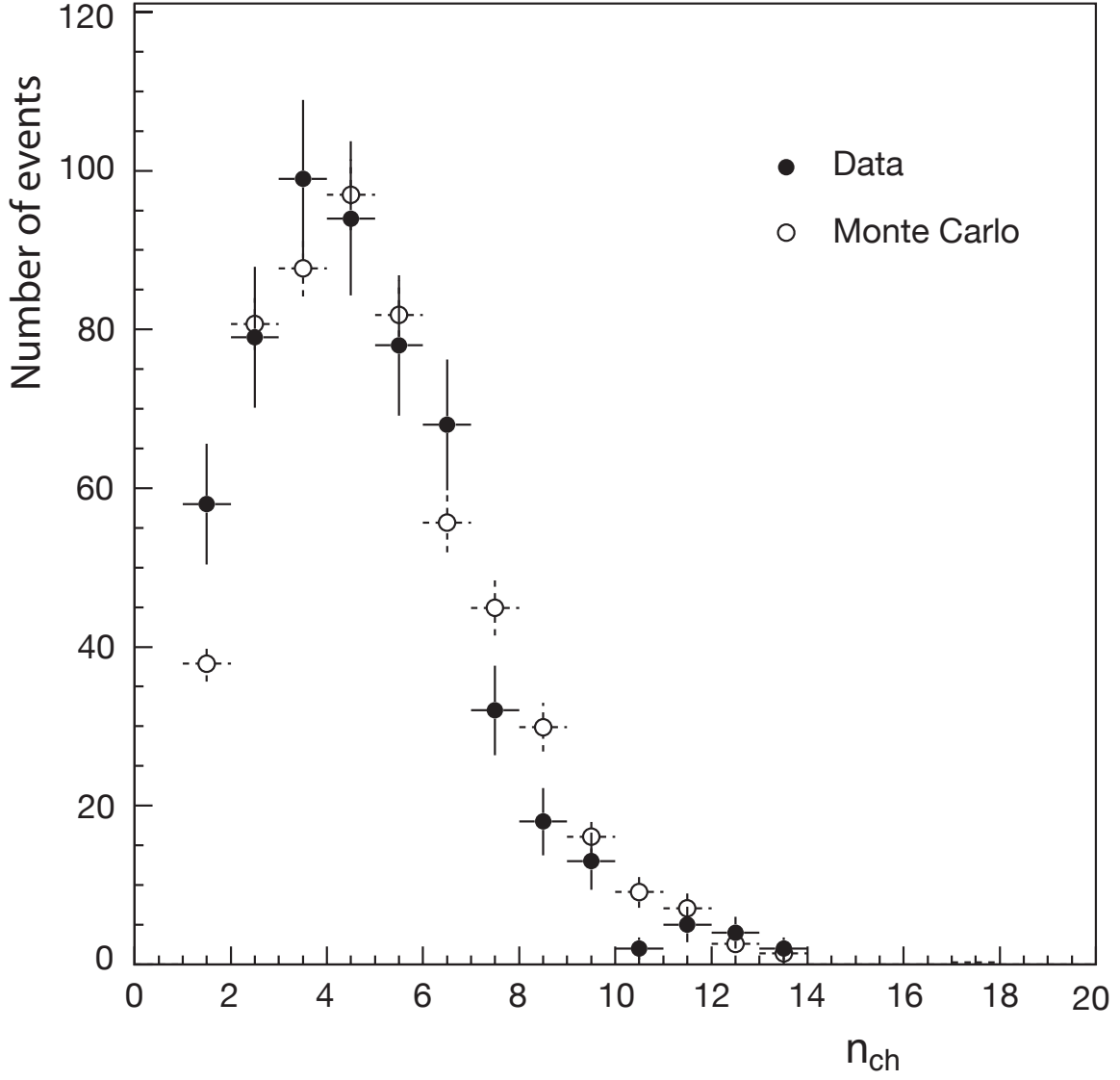}
\caption{Charged-particle multiplicity, $n_{ch}$, at the primary vertex
of all the located events. Data is shown by solid circles, Monte Carlo by open circles.}
\label{fig:mult}
\end{figure}

We investigated the located-event sample for possible biases. Fig. \ref{fig:z} displays the distance along the beam direction between the vertex and the downstream edge of an emulsion module, for all 7 modules. The distribution is consistent with being independent of $z$, with $\chi^2$/ndf to a straight line of 1.7. The vertex distribution in the transverse plane (not shown) is uniform, as expected. The located-event charged multiplicity distribution is compared with expectation in Fig. \ref{fig:mult}. We conclude that the benefit of using a combination of different location methods was to have uniform location
efficiency.

\section{Secondary vertex analysis}
\label{sec:decay-vtx}

\subsection{Decay search criteria}
\label{sec:decay-crit}

For the located events, the emulsion was digitized again in a smaller
volume  containing the vertex and optimized for the decay search, typically
$2.5 \rm{mm} \times 2.5 \rm{mm} \times 12 \rm{mm}$.
The track reconstruction algorithm was the same as that used for
 vertex location. The decay search was divided into
 two categories distinguished by topology:
\begin{enumerate}
\item {\it Long-decay search:} Decays in which the candidate parent track
 passed through at least one emulsion layer.
\item {\it Short-decay search:} Decays in which only the daughter track
was recorded in emulsion.
\end{enumerate}
The strategy was common for both decay topologies under consideration.
Once a secondary vertex was found, the event was classified as a one-prong
decay, unless additional tracks were found to be associated with the same
secondary vertex constituting a three-prong decay.

Tau and charm decays were obtained from the data in a
two-step process: {\it (i)} finding secondary vertices in emulsion data using
geometrical cuts, described in this Section, 
 and {\it (ii)} subsequently imposing topological and kinematical 
cuts to isolate the signal from the background, described in
 Section \ref{sec:recog}.

\subsubsection{Long-decay search}
\label{sec:Long}

The Long-decay search for one-prong decays imposed the following criteria:
\begin{itemize}
\item  The parent track had one or more microtracks, and a daughter track had
 three or more microtracks.
\item   The parent track length: $L_{dec}$ $<$ 10 mm.
\item  The  impact parameter $b_p$ of the parent track
 with respect to the primary vertex:
 ({\it i})  $b_p < 5\ \mu$m if there were at least two microtracks, or
({\it ii}) $b_p < (5 + 0.01\times \delta z$) $\mu$m if there was
 one microtrack, where $\delta z$ is the distance from the
 parent microtrack to the vertex.
\item  The minimum distance, $d_{min}$, between extrapolated parent and
 daughter tracks:
({\it i}) $d_{min} < 5$ $\mu$m if there were at least two parent microtracks,
 or ({\it ii}) $d_{min} < (5 + 0.01 \times \delta z)$ $\mu$m if there was
 only one parent microtrack.
\item ({\it i}) The angle between the daughter and parent tracks:
   $\alpha >$ 4 times the angular measurement error, or
 ({\it ii}) The impact parameter $b_d$ of the daughter with respect to
 the primary vertex: $b_d >$ 4 times the error in the position.
\end{itemize}

Candidate tracks passing the above criteria were checked in the emulsion
by a physicist using a microscope to ensure that (\emph{i}) the daughter track
could not be associated with emulsion tracks upstream of the vertex, (\emph{ii})
that it was not a part of a e$^+$e$^-$ pair, and (\emph{iii})
that there were no alignment problems.

\subsubsection{Short-decay search}

The Short-decay search for one-prong decays required the following criteria:

\begin{itemize}
\item The daughter track had at least three microtracks.
\item The daughter-track impact parameter (IP) with respect to
 the primary vertex: $b_d < 200$ $\mu$m.
\item The daughter-track IP w.r.t. the primary vertex:
 $b_d > 4 \times \sigma_{IP}$, where $\sigma_{IP}$ is
 the error on the impact parameter.
\end{itemize}

Each candidate daughter track was checked visually to insure
 that it could not be connected to microtracks upstream of the vertex.

\subsection{Tau and charm recognition}
\label{sec:recog}

To extract the $\nu_\tau$ signal from events passing the secondary-vertex
selection, a set of topological and kinematical criteria was first applied
 as described in Section \ref{sec:recog-topo} below.
In the second step, the amount of signal and background was determined using
a multivariate technique featured in Section \ref{sec:multivar}.
 
\subsubsection{Topology and kinematical cuts}
\label{sec:recog-topo}

{\it $\nu_\tau$ event topology.} The $\nu_\tau$ CC interactions,
 reaction (\ref{eq:tauCC}), produce a $\tau$ lepton
 that typically decays within 2 mm of its origin. Thus,
 the topological signature for $\nu_\tau$ events is a track
 from the primary vertex that gives a secondary vertex
 at a short distance consistent with the kinematics
 of the decay. There must be no other
 lepton from the primary vertex. 
 The topological signature of charm production in reaction (\ref{eq:charm})
 is very similar to $\nu_\tau$ events.
 Tau and charm events were distinguished primarily by presence of
 an electron or muon at the interaction vertex. Thus, a $\nu_e$ or a $\nu_\mu$
 CC interaction together with a failure in lepton identification constitutes
 the primary background to the tau sample.
The second background considered here were interactions of hadrons produced in
neutrino NC interactions, reaction (\ref{eq:scatter}), that appeared in the
emulsion with a topology of  a one-prong or three-prong interaction (or decay).
 
{\it Kinematical cuts.} The following set of criteria were derived from
 Monte Carlo studies to efficiently extract the $\nu_\tau$
 signal with minimal background. It is a modified
 version of the selection criteria of  Ref. \cite{ref:DONuTfirst}.
 Long one-prong and trident decays were accepted when
 the following conditions were satisfied:

\begin{itemize}
\item  Parent-track angle w.r.t. neutrino direction: $\theta_p < 0.2$ rad.
\item  Daughter-track angle w.r.t. parent direction: $\theta_d < 0.3$ rad.
\item  Kink angle:  $\alpha < 0.25$ rad.
\item  Daughter-track IP:  $b_d < 500$ $\mu$m.
\item  Transverse momentum of the daughter w.r.t. parent track:
 $p_T > 250$ MeV/$c$ for hadrons,
 and $p_T > 100$ MeV/$c$ for electrons and muons.
\item  Daughter momentum: $p_d > 1$ GeV/$c$.
\end{itemize}

\begin{figure}
\includegraphics[width=\columnwidth]{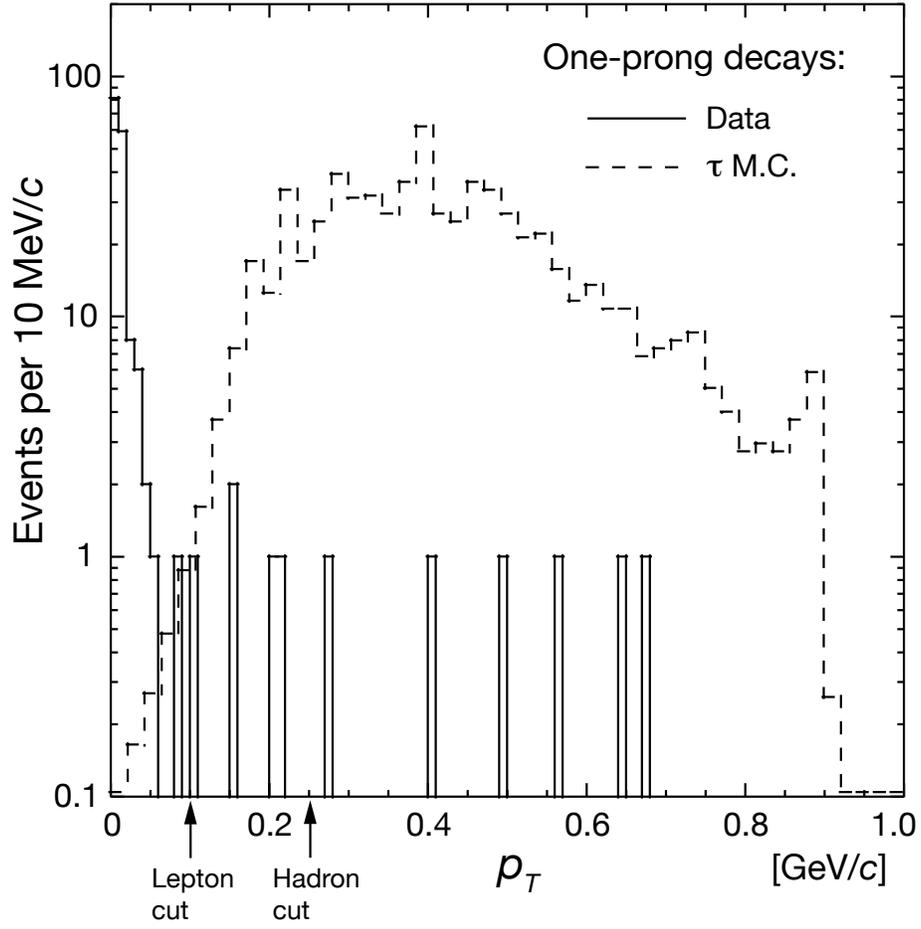}
\caption{
The distributions of one-prong secondary-vertex events (solid line)
 after all the topological and kinematic cuts {\it except} on transverse momentum.
 Superposed is the expected distribution from $\tau$ one-prong decays (dashed line, arbitrary
 normalization).
 For $\tau$ candidates, the kink transverse momentum  must exceed 0.25 GeV/$c$
for $\tau \rightarrow hadron$ or exceed 0.1 GeV/$c$ for $\tau \rightarrow$ $e$ or $\mu$. }
\label{fig:pt_kinks}
\end{figure}

Events passing these criteria that did not have an identified
 electron or muon track from the primary interaction vertex
 were selected as $\nu_\tau$ candidate events.
In the case of trident secondary vertices, at least one
 of the secondary tracks must pass all of the above requirements. 
Fig.\ \ref{fig:pt_kinks} shows the distribution of number of kinks versus 
transverse momentum,  $p_T$, of the daughter w.r.t the parent track,
 for all tracks satisfying the above criteria \emph{except}
 the transverse momentum cut.
 One can see that $p_T$ is an impressive discriminant.
 There are 198 tracks, but  almost all are within the steeply falling
peak  at low $p_T$ due to hadronic background, reaction (\ref{eq:scatter}).
All but one of the other tracks are classified as either tau or
charm decays following the multivariate analysis outlined in the next section.

For Short decays, all the cuts were the same but one: the kink angle $\alpha$
cannot be defined since the parent direction is unknown. 
 Here the kink angle was replaced by the ``minimum kink angle'',
obtained by extrapolating the daughter track back to the steel plate and
 placing the ``decay vertex" at the point where this extrapolation intersects
 the downstream face of the plate.
This was the most conservative assumption, since it
 also minimized the transverse momentum assigned to the decay.

\subsubsection{Mutivariate analysis}
\label{sec:multivar}

Only events selected by secondary vertex analysis detailed above
 were submitted to the multivariate analysis
 employed to determine the probability that
individual events represented one of the following interaction types,
 each with a one-prong or a three-prong secondary vertex:
\begin{enumerate}
\item $\nu_\tau$ CC events, reaction (\ref{eq:tauCC}).
\item Charm production, reaction (\ref{eq:charm}).
\item  Neutrino NC events with a secondary hadron interaction,
 reaction (\ref{eq:scatter}).
\end{enumerate}
No other physical process, subject to the topological
 and kinematical cuts above,
 was deemed to be a significant part of the background.

A set of quantities was chosen that could be easily and unambiguously
 measured in the emulsion data (supplemented by spectrometer information)
 and that could discriminate between the three hypotheses.
Note that all these quantities are independent of the
neutrino production and interaction processes.
 For $n$ parameters, an $n$-dimensional probability density distribution
 for each hypothesis was computed using Monte Carlo generated events.
 Then the relative probability of event $k$ sampled from the distribution
 of hypothesis $i$ can be written as
\begin{equation}
P(\{x_k\}|i) = \frac{{\mathcal W}_i{\mathcal P}(\{x_k\}|i)}
{\sum\limits_j{\mathcal W}_j{\mathcal P}(\{x_k\}|j)}
\label{eq:bayes}
\end{equation}
where $\{x_k\}$ is a set of parameters describing event $k$,
 ${\mathcal P}(\{x_k\}|i)$ is the probability density function
for hypothesis $i$ evaluated for $x_k$ determined from the data,
 and ${\mathcal W}_i$ is the prior probability of the event
 being an $i$-type event. 
Note that the ${\mathcal W}_i$ are independent of $\{x_k\}$,
and give the probability of a neutrino interaction of type $i$
 occurring within the emulsion fiducial volume using full MC simulation
  starting with neutrino production in the beamdump
 through its interaction in the emulsion target.

The parameter set $\{x_k\}$ for events selected as tau
candidates included $L_{dec}$, $\alpha$, $p_d$,  $\theta_p$, and  $\sum b_d$,
introduced above. In addition,  $\Delta \phi$ was added, which represents the
 angle in the plane transverse to the neutrino beam
 between the parent direction and the vector sum of unit vectors of the
 remaining tracks at the primary vertex, expected to peak at 180$^{\circ}$
for $\nu_\tau$ CC events, and to distribute uniformly for the other two
hypotheses.

Hence, for one-prong decay candidates resulting from the Long-decay search,
 the set $\{x\} = \{L_{dec},\ \alpha,\ p_d,\ \theta_p,\
 \Delta \phi\}$ was used,
and $\{x\} = \{L_{dec},\ \theta_p,\ \Delta \phi,\ \sum b_d\}$ 
was used for three-prong decays.

\begin{figure}
\includegraphics[width=\columnwidth]{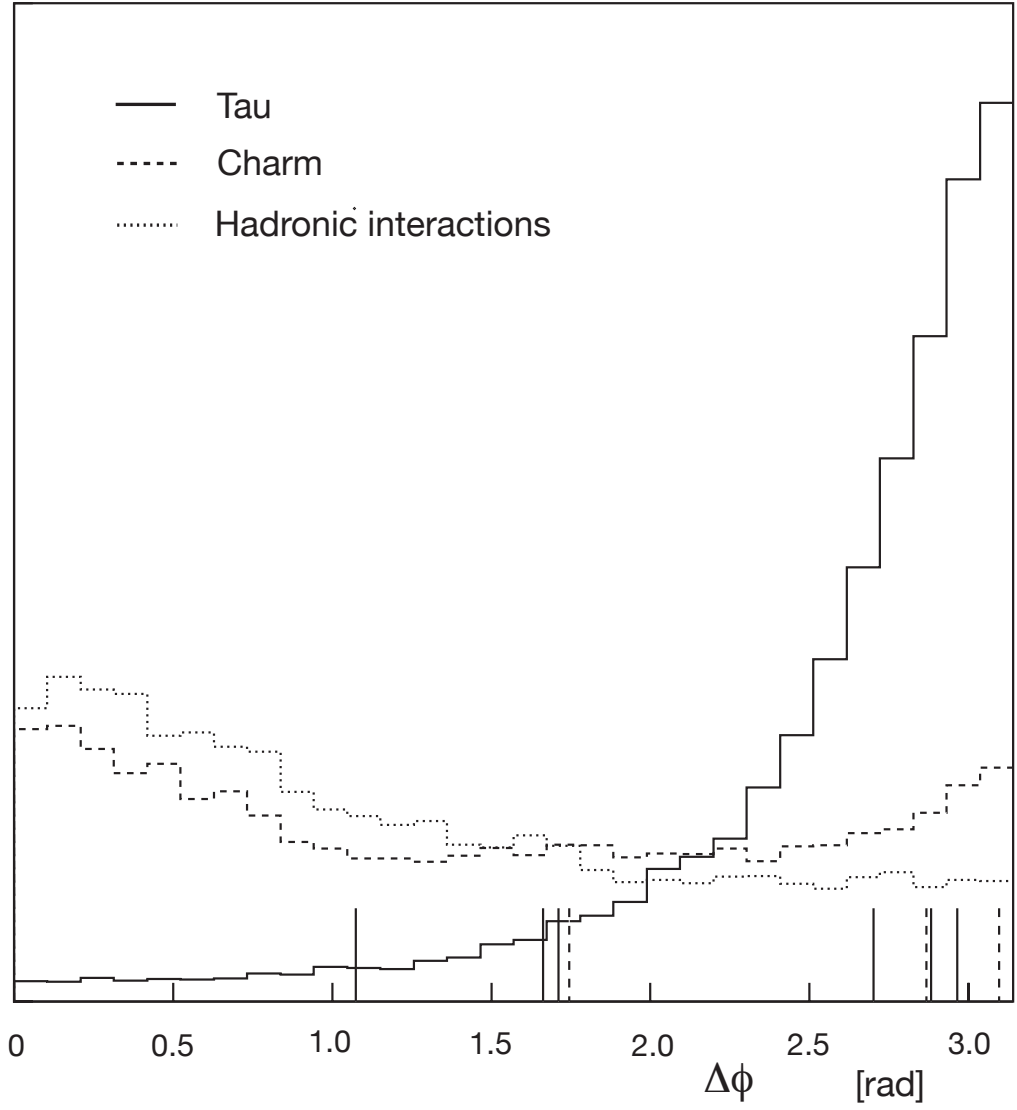}
\caption{
An example of simulated distributions used as input to the event probability calculation within the multivariate method as applied to all decays.
Shown are distributions of the transverse-plane angle $\Delta\phi$
for all three hypotheses under consideration:
tau ({\it solid line}), charm ({\it dashed line}), and hadronic interactions ({\it dotted line}).
Short vertical lines indicate the values for $\nu_\tau$ candidate events from 
Table \ref{tbl:nutau} for one-prong
decays ({\it solid line}) and three-prong decays ({\it dashed line})}
\label{fig:compare_phi}
\end{figure}

\begin{figure}
\includegraphics[width=\columnwidth]{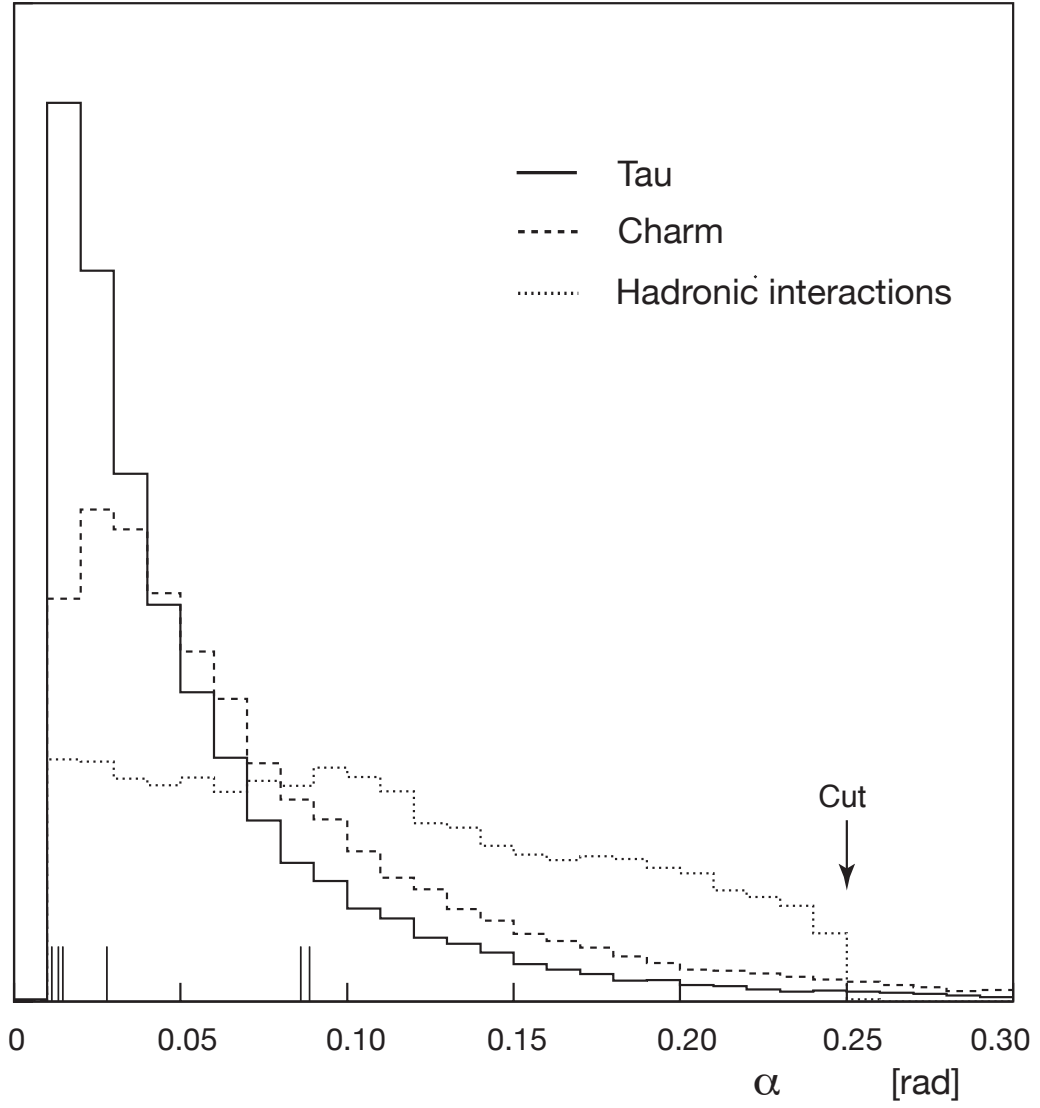}
\caption{
An example of simulated distributions used as input to the event probability calculation within the multivariate method as applied to one-prong decays.
Shown are distributions of the kink angle $\alpha$ for all three hypotheses under consideration:
tau ({\it solid line}), charm ({\it dashed line}), and hadronic interactions ({\it dotted line}).
Short vertical lines indicate the values for $\nu_\tau$ candidate events from 
Table \ref{tbl:nutau}.}
\label{fig:compare_alpha}
\end{figure}

\begin{figure}
\includegraphics[width=\columnwidth]{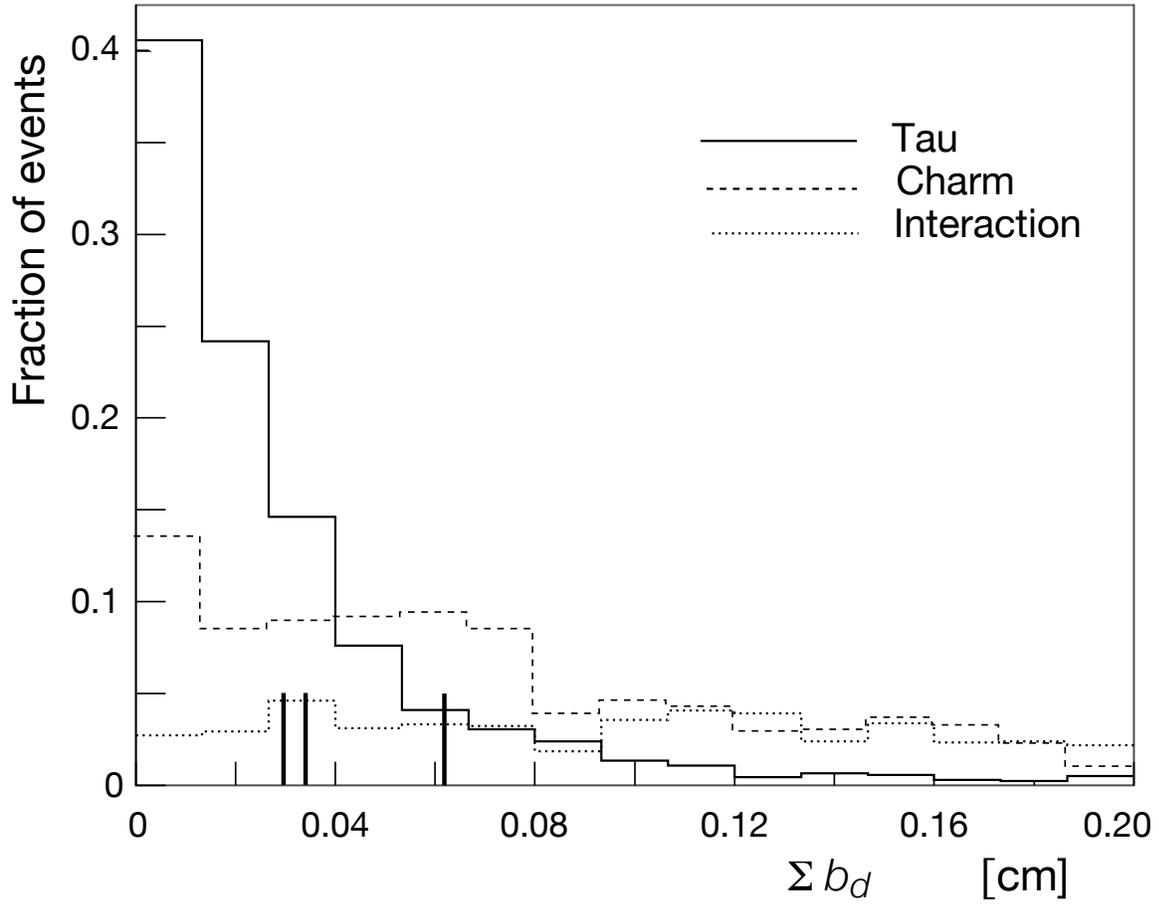}
\caption{
An example of simulated distributions used as input to the event probability calculation within the multivariate method as applied to three-prong decays.
Shown are distributions of $\sum b_d$, the sum of daughter-track impact parameters
for all three hypotheses under consideration:
tau ({\it solid line}), charm ({\it dashed line}), and hadronic interactions ({\it dotted line}).
Short vertical lines indicate the values for $\nu_\tau$ candidate events from 
Table \ref{tbl:nutau}.
}
\label{fig:compare_SumIP}
\end{figure}

 Simulated distributions used as
input to the multivariate method are illustrated in
 Figures \ref{fig:compare_phi}-\ref{fig:compare_SumIP} for all three hypotheses.
Fig. \ref{fig:compare_phi} shows the $\Delta\phi$ angle in the transverse plane, 
used for both  one- and three-prong topologies, which discriminates very strongly
against both charm and hadronic-interaction background.
 Fig. \ref{fig:compare_alpha} shows the $\alpha$ decay angle  
used for  the one-prong topology, which discriminates strongly
against the hadronic-interaction background, and provides 
modest discrimination against charm.
Fig. \ref{fig:compare_SumIP} shows $\sum b_d$, sum of
the daughter-track impact parameters, used for the three-prong
topology. This quantity is related to $ct$ for this event, where $t$ is
this parent's lifetime in its rest frame. Since $\tau$-lepton has shorter
lifetime than charmed mesons,
 $\sum b_d$ discriminates very strongly against the hadronic-interaction
background, and provides strong discrimination against charm.
Note that these one-dimensional distributions do not provide
information about correlations among the multivariate parameters
which are taken into account in the calculation.

 The multivariate analysis was also used for events from the Short-decay
 search. Here, the parent direction is
 unknown, and hence $\theta_p$,  $\alpha$ and $\Delta \phi$ are unkown.
 The true decay point must have been in the same steel plate
 that contained the interaction vertex, lying
 on a line  made by projecting the candidate daughter track upstream.
 Along this line within the steel, the parameters
 $L_{dec}$, $\alpha$, $\theta_p$, and $\Delta \phi$ vary continuously, so that
 probabilities for the three hypotheses also vary.
 To make a definite and conservative estimate, the values of all three
 probabilities were measured at the point along the line where
 the tau-hypothesis probability was minimum.

Table \ref{tbl:priors} summarizes the prior probabilities for both kink
 and trident topologies and different materials of the emulsion target.
Resulting hypothesis probabilities for the $\nu_\tau$ event candidates
are presented in Section \ref{sec:nutau-sig} below.

\subsection{Decay search efficiencies}
\label{sec:DecayEff}

The effect of cuts applied during the secondary vertex search was determined by
Monte Carlo calculation for all three hypotheses, tau, charm, and
hadronic interaction.
The secondary-vertex search efficiency was checked by using secondary hadronic interactions
found as a byproduct of the track-by-track electron ID scans. 
The number of interactions expected has a well-understood value depending on path length
in a given material (emulsion, steel or plastic).
The number of interaction vertices of all multiplicities was estimated to be 31.
The total number of found interactions was 27, yielding an
efficiency of 0.87,  consistent with a Monte Carlo derived
efficiency of 0.86.

The fractions of events remaining after selections described in Secions \ref{sec:decay-crit} and
\ref{sec:recog-topo} are listed in Table \ref{tbl:taueff}. The estimate for the overall systematic
 uncertainties in these efficiencies is 5\% of the value.

\section{Survey of data}
\label{sec:data-survey}

\subsection{Expected composition}
 \label{sec:expect}

The expected number of interactions for
 reactions (\ref{eq:eCC}) - (\ref{eq:NC})
 was predicted using the DONuT Monte Carlo simulating the same event-selection
 procedure that was applied to the data. Charged-current ineractions of
 all flavors were selected by identifying a lepton at the primary vertex.
 All neutrino interactions without an identified lepton were considered
 to be ``effective neutral-current" events, NC$_{\rm{{\it eff}}}$.
 These NC$_{\rm{{\it eff}}}$ events therefore included
 CC events with a lepton that escaped detection.
 Table \ref{tbl:expect} shows the expected number of events of all four
 interaction types. Note that although the prompt and non-prompt components
 (see Section \ref{sec:beamdet}) are separated in the simulation, they are
 not distinguishable in the data.

\subsection{$\nu_\mu$ CC events}
\label{sec:numuCC}

The identification of muons using the spectrometer was
 straightforward and efficient, so this category of
 interactions was considered the most reliable.
 The number of $\nu_\mu$ CC events found was 225 events, which
 gives the fraction of  $\nu_\mu$ CC to the total (578) as $0.39 \pm 0.03$.

The fraction of prompt $\nu_\mu$ CC events was estimated both by Monte Carlo and from the data.
Averaging over several algorithms, the MC estimate is $0.61\pm0.03$. An estimate from
combining results from analyses based on data (number of $\nu_e$ CC interactions, fitting
to the muon spectrum and data taken with a half-density beamdump) gives $0.59\pm0.06$.
The estimated number of prompt $\nu_\mu$ CC interactions is thus $133\pm16$.
 
The ratio of the number of  $\overline{\nu}_\mu$ interactions with outgoing $\mu^+$
 to the number of $\nu_\mu$ interactions with $\mu^-$ was computed
 from $\nu_\mu$ and  $\overline{\nu}_\mu$ cross sections taking into account
 detector efficiency and acceptance. The resulting expected ratio was 0.63.
 The same fraction determined in the 578-event data sample was $0.67 \pm 0.08$.
 Using this measured ratio, the ratio  of  integrated  $\overline{\nu}_\mu$ and
 $\nu_\mu$ fluxes was found to be $1.05 \pm 0.13$.
 
 There are three events in the located sample that have two
 identified muons. One event has muons of opposite sign with
 one from the primary interaction vertex and the other from
 a secondary decay vertex. This event is identified as
 a $\nu_\mu$ CC interaction producing a charmed meson. The other
 two dimuon events have same-sign tracks, where one of the tracks is likely
 a charged $\pi$ decaying in-flight.

\subsection{$\nu_e$ CC events}
\label{sec:nue}
 
 The expected mean energy of outgoing electrons in $\nu_e$ CC interactions
 was 52 GeV, with 22\% of events having electron energies below 20 GeV. 
 Approximately 15\% of NC events have at least one electron with 
 energy less than 20 GeV.  Therefore, a low-energy cut is
 applied to the electron sample to reduce background from events that
 are not $\nu_e$ CC events. Table \ref{tbl:electron_ID} summarizes the
 result of a Monte-Carlo-based study to optimize this cut and to estimate the
 NC background as a function of energy. For cuts of 18 GeV and higher, there
 is little change in signal-to-background ratio and a cut of 20 GeV was chosen.
 A total of 120 $\nu_e$ CC and NC$_{\rm{{\it eff}}}$ events passed the cut.
The NC$_{\rm{{\it eff}}}$ background fraction is estimated in Table \ref{tbl:electron_ID}
 to be 0.174, so the best estimate for the number of $\nu_e$ CC events
 (with a 20-GeV electron cut) is given as $120 \times (1-0.174) = 99 \pm 9$,
as determined by the electronic detector data. To compare this number to
the second identification method which follows, it must be divided by the
electronic identification efficiency (0.80), yielding $124\pm11$. 

 The set of events with electrons identified in the emulsion was analyzed
 independently. There were 82 events with primary electrons found in the
 emulsion data alone. Of these, 62 electrons passed the 20 GeV minimum
 energy cut. The electron-identification efficiency of this procedure
 was found to be independent of energy. The number of $\nu_e$ CC,
 corrected by the efficiency, was $62/0.66 = 94 \pm 12$.

\subsection{$\nu_\tau$ CC events}
\label{sec:nutau-sig}

The methods of selecting the $\nu_\tau$ events described in
 Section \ref{sec:decay-vtx} were applied to the 578 located events.
The multivariate analysis (Section \ref{sec:multivar}) was performed
for each selected event.  Events with  $P(\tau) > 0.5$ are listed in Table \ref{tbl:nutau}. 
We estimate the number of $\nu_\tau$, charm, and hadronic interaction events
 in our final sample by summing up the hypothesis probabilities in 
Table  \ref{tbl:nutau}, yielding 7.5 $\nu_\tau$ events, 1.26 charm events,
and 0.22 hadronic interactions.

The charm and hadronic-interaction backgrounds can also be estimated
in the tau sample using one-dimensional cuts on Monte Carlo events
without any reference to the correlations between variables. This
simpler analysis gives an estimate of the background from charm decays and
hadronic interactions in the nine selected events as 1.1 and 0.9 events,
 respectively. In comparing the results between the two analyses, it is
 important to note that the multivariate method accounts for correlations
 between parameters and results depend on the particular set of candidate
events. This last point is significant due to the small number of tau events.
The similarity of the charm background from the two analyses demonstrates
the similarity in the topological signature of tau and charm decays.
The hadronic interaction background, however, shows little correlation
 between parent track length and `decay' (interaction) topology, and simple
 one-dimensional cuts overestimate this background.

 \subsection{Charm production in neutrino interactions}
 

Integrating over the expected neutrino energy spectrum,
 the average charm production fraction,
 normalized to the number of $\nu_\mu$ and $\nu_e$ CC interactions, is  
$0.066 \pm 0.008$ \cite{ref:jason}.
 This fraction includes production of D$^0$, 
D$^\pm$, D$_s$, and $\Lambda_c$. Including only charged charmed 
hadrons reduces the fraction to $0.028 \pm 0.006$. 
The expected number of charged charm events is the product
 of the total number of located events (578),
the fraction of CC events (0.62), the efficiency for observing the
secondary decay ($0.45 \pm 0.05$) and the charged charm fraction (0.028). 
The result is $4.5  \pm 1.0$ events, where the error represents
 the uncertainties in cross sections and branching ratios.
The observed number of charged charm events in 
our sample is 7 events, with an estimated background level of 2.2 events,
which is consistent with our prediction.

\section{Nu-tau cross section}
\label{sec:cross_sec}
\subsection{Analyses}

Two methods were used to measure the cross section for 
$\nu_\tau$-nucleon CC interactions. The primary analysis
 determined the ratio of the $\nu_\tau$-nucleon cross section
and the $\nu_e$-nucleon or $\nu_\mu$-nucleon cross section.
Systematic uncertainties in neutrino production that affected
all  flavors equally canceled in the relative measurement.
Electronic triggering efficiencies and neutrino interaction
selection  efficiencies were high and
for CC events showed no dependence on flavor.
However, some corrections applied to the data did not cancel,
and their uncertainties contribute to the systematic error.
Since only the prompt $\nu_\mu$ are relevant in the relative
cross section  calculation, the uncertainty in the prompt fraction
was included in the  systematic error of the
$\sigma(\nu_\tau N)$ to $\sigma(\nu_\mu N)$ ratio.
Similarly, the systematic uncertainty related to the energy cut
and the NC background subtraction in the $\nu_e$ sample
 was included in the $\sigma(\nu_\tau N)$ to $\sigma(\nu_e N)$ ratio.
The $\nu_\tau$ analysis required the secondary vertex search, and
this efficiency (0.46) is applied to the $\nu_\tau$ events.

The second technique measured the absolute cross section for
$\nu_\tau$N CC interactions. All electronic, event selection and
analysis efficiencies appear  explicitly in the calculation.

The cross section calculations required an estimate for the number of
neutrino interactions in the emulsion target, corrected for efficiency and  acceptance.
It was important to account for correlations between acceptance and  energy.
The number of interactions of each flavor in the experiment can be  written as

\begin{equation}
\label{eq:no_int}
N_{{\rm{int}}}  = \frac{{N_\nu ^{{\rm{tgt}}} }}{{N_{\rm{pot}}}} \cdot  N_{\rm{pot}} \cdot {\varepsilon} \cdot
\frac{{\sigma ^{{\rm{const}}} }}{{Area}} \cdot \frac {{M_{{\rm{tgt}}} }}{{m_{{\rm{nucleon}}}
}} \cdot \frac{f}{{N_\nu ^{{\rm {MC}}} }}\sum {EKTt\,}  ={\sigma^{\rm{const}}} {\varepsilon} C f\left\langle {\sum {EKTt} }
\right\rangle
\end{equation}
where the sum is over neutrinos generated by Monte Carlo
in the beamdump  with energy $E$, and with kinematic
suppression factor $K(E)$.  The binary $T$ was  equal to one
if the neutrino passed within the target fiducial volume and
the  binary $t$ was equal to one if the interaction generated a trigger.
The number of neutrinos generated in the Monte Carlo is denoted by
$N_ \nu^{\rm{MC}}$ and the number
passing through the emulsion is $N_\nu ^{\rm{tgt}}$.
The area is taken to be the size of the emulsion, 50 cm $\times$ 50 cm.
For the Monte Carlo events, the simulated trigger also incorporated  the muon
identification for $\nu_\mu$ interactions. 
The electron  identification, with its
efficiencies, was not incorporated directly  into $t$ but it, as well as 
other electronic and analysis efficiencies, were incorporated into $\varepsilon$.
The number of protons accumulated in the beamdump ($N_{\rm{pot}}$) and the 
fraction of  the neutrino flux ($f$) intercepting the emulsion are
explicitly shown. The quantity $C$ incorporates the energy-independent factors
and it depends on neutrino flavor. The angle brackets indicate that the
mean value  of the sum of the products is used.
The Monte Carlo gives $f$ and the  mean value of the sum directly and
the constants of Eq.\eqref{eq:no_int} are incorporated  into $C$. The values of $C$ and the sum
that were used in this analysis are  listed in Table \ref{tbl:A1}.

The total CC cross sections per nucleon can be written
\begin{equation}
\sigma _{\nu _\ell  }  = \sigma _{\nu _\ell }^{{\rm{const}}} \,E\;K\left ( E \right),
\hspace{6mm}\ell = e,\mu,\tau
\label{eq:CS4}
\end{equation}
where $\sigma _{\nu _\ell }^{{\rm{const}}} $
is the energy-independent factor of the cross section of flavor $\ell$,
and $K$ gives the part of the tau-neutrino cross section that depends
on kinematic effects due to the $\tau$-lepton mass (see Fig. \ref{fig:kinfac}).
In the DONuT energy range (Fig. \ref{fig:nu_spectrum}),
the factor $K$ can be safely taken to be unity for $\nu_e$ and
$\nu_\mu$ CC interactions.
With this notation, the relative cross  sections can be written,
\begin{equation}
\frac{{\sigma _\tau ^{{\rm{const}}} }}{{\sigma _\ell^{{\rm{const}}} }} =  \frac{{N_\tau
^{{\rm{exp}}} }}{{N_i^{{\rm{exp}}} }} \cdot \frac {{C_\ell }}{{C_\tau  }} \cdot \frac{{f_\ell
\left\langle {\sum {ETt} }  \right\rangle _\ell }}{{f_\tau  \left\langle {\sum {EKTt} }
\right \rangle _\tau  }} \cdot \frac{{\varepsilon _\ell }}{{\varepsilon _\tau  }},
\hspace{6mm}\ell = e,\mu
\label{eq:CS5}
\end{equation}
The  $\varepsilon_i$ denote efficiencies for lepton identification  only.
The efficiency of the secondary vertex search is included in $ \varepsilon_\tau$.

The absolute $\nu_\tau$ cross section is computed from the following  expression,
\begin{equation}
\sigma _\tau ^{{\rm{const}}}  = \frac{{N_\tau ^{\exp } }} {{\varepsilon _{{\rm{TOT}}} 
\cdot C_\tau   \cdot \left( {f\left \langle {\sum {EKTt} } \right\rangle } \right)_\tau 
}}
\label{eq:abs_cs}
\end{equation}
where $\varepsilon_{\rm{TOT}}$ is the product of all experimental  efficiencies
\begin{equation}
\varepsilon_{\rm{TOT}} = \varepsilon_{\rm{FS}}\cdot\varepsilon_{\rm {trig}}\cdot
\varepsilon_{\rm{loc}}\cdot\varepsilon_{\rm{\tau}}.
\label{eq:eff_tot}
\end{equation}
The efficiencies in Eq. \eqref{eq:eff_tot} are as follows:
 filtering and  scanning ($0.85\pm0.06$), trigger with live-time ($0.79\pm0.02$),
location in emulsion ($0.64\pm0.04$), and secondary vertex finding ($0.46\pm0.02$),
yielding  $\varepsilon_{\rm{TOT}}$ = $0.20 \pm 0.02$.

\subsection{Systematic uncertainties}

The  cross section results from this experiment depend on predicting the
neutrino fluxes of each flavor. 
The value of $C_\ell$ in Eq. \eqref{eq:CS5} and Eq. \eqref{eq:abs_cs} depends linearly on the
total charm production cross section in $pN$ interactions in the beamdump. 
And the value of $f_\ell$ times the term in the brackets depends on the angular distribution 
of charm in the $pN$ center-of-momentum frame. Most of the systematic uncertainty in the cross section
results was due to these two terms. We examine each in more detail. 

The factor $C_\ell$ contains the number of neutrinos produced in the beamdump, so it is
sensitive to variations in total cross section, branching ratios and target atomic number effects, 
which we parameterize by  $A^{\alpha}$. 
The relative errors for charm production of $\nu_e$ and $\nu_\mu$
is taken to be the same for both: 0.10 from charm total cross section, 0.16 from branching ratios and 0.14 from the $A$ dependence.
We adopt the convention to add the errors in quadrature where values are derived from several
sources and not likely to be correlated. 
This gives a total relative error of 0.23 for $C_e$ and $C_{\mu}$.
The estimated uncertainty in $C_{\tau}$ depends almost entirely on $\rm{D}_s$ production and
decay. The relative uncertainties are computed to be 0.15, 0.23 and 0.14 for cross section, branching
ratio and $A$ dependence, respectively. Added in quadrature, this gives 0.31 for the relative 
uncertainty in $C_{\tau}$.
In the results for the relative cross section measurement, below, the uncertainty in the $A$ 
dependence is not included in the second, systematic error.

The factor $f\Sigma EKTt$ is sensitive to kinematic uncertainties in charm production, with
the effects manifested in the variation of the parameter $n$ of Eq. \eqref{eq:charm_cs}. Both the
neutrino energy (and hence number of interactions) and the fraction of the neutrino flux 
within the emulsion are affected. We compute the amount of variation in the number of
accepted Monte Carlo events and assign it to the systematic error in $f\Sigma EKTt$. 
We assume $n = 8.0\pm0.8$ for both $\rm{D}_s^+$ and $\rm{D}_s^-$ production, but in
computing the relative error, allow $n$ to be different by $\pm2.0$ for $\rm{D}_s^-$.
This gives a relative uncertainty of +0.31 and -0.23 in $\nu_{\tau}$ production. The uncertainties
in $f\Sigma EKTt$ for $\nu_e$ and $\nu_\mu$ were computed analogously, yielding +0.30 and
-0.20. The positive uncertainty corresponds to a decrease in $n$ by two units.

For $\nu_e$ and $\nu_\mu$ CC interactions we can estimate $C_{e,\mu}$ from the number of 
interactions in the data, given the values of $f\Sigma EKTt$ and the efficiencies
computed from the Monte Carlo. This provides a systematic
check on $C$. The values are $C_e = 1.47\times10^{40}\rm{ cm}^{-2}$ and $C_{\mu} = 1.79\times10^{40}\rm{ cm}^{-2}$ (prompt muons only).
These are compared with $1.64\times10^{40}$ and $1.55\times10^{40}$, respectively from Table \ref{tbl:A1}, which were 
extracted from Monte Carlo
simulations with values of the parameter $n$ discussed above.
This indicates that the
systematic uncertainty in the charm cross sections is within the values (+0.30, -0.20) estimated above.

\subsection{Results}

The relative cross sections were obtained from Eq. \eqref{eq:CS5} using the observed number of interactions,
corrected by efficiency and kinematic factors.
Inserting the values from Table \ref{tbl:cs_factors} yields
\begin{equation}
\frac{{\sigma _{\nu _\tau  }^{{\rm{const}}} }}{{\sigma _{\nu _e }^ {{\rm{const}}} }} =
1.58 \pm 0.58 \pm0.91 \quad {\rm{and}}\quad \frac{{\sigma _ {\nu _\tau  }^{{\rm{const}}}
}}{{\sigma _{\nu _\mu  }^{{\rm {const}}} }} = 1.16 \pm 0.42\pm0.65
\label{eq:rel_cs-result}
\end{equation}

The first error in the results is the statistical error, the second is the estimated sytematic
uncertainty. The systematics of these two results are correlated, since the same assumptions
regarding charm production were made for both $\nu_e$ and $\nu_\mu$ production. 
Therefore, the two cross section may be averaged without introducing other uncertainties.
The result is 

\begin{equation}
\frac{{\sigma _{\nu _\tau  }^{{\rm{const}}} }}{{\sigma _{\nu _{e,\mu}}^ {{\rm{const}}} }} =
1.37 \pm 0.35 \pm0.77
\label{eq:rel_cs-result_avg}
\end{equation}

The absolute $\nu_\tau$-nucleon cross section was computed using the  factors of Table
\ref{tbl:A1} inserted into Eq. \eqref{eq:abs_cs}:
\begin{equation}
\sigma _{\nu _\tau}^{{\rm{const}}}  = 0.72 \pm 0.24 
 \pm 0.36\times 10^{-38} \quad \rm{cm}^2\,{\rm{ GeV}}^{{\rm{-1}}}
\label{eq:abs_cs-result}
\end{equation}
The first error is statistical, the second one systematic.

Lack of knowledge of the charge of the $\tau$ lepton
implies that the result, Eq. \eqref{eq:abs_cs-result},
represents an average of $\nu_\tau$ and $\bar \nu _\tau$ cross sections.
The measured value of  $\sigma _{\nu _\tau}^{{\rm{const}}}$
is to be compared with the average of $\nu_\mu$ and
$\bar \nu_\mu$ cross section factors,  $0.51 \times 10^{-38}$ cm$^2$ \cite{ref:PDG}, 
assuming equal fluxes of neutrinos and antineutrinos
in the DONuT beam. Hence, the $\nu_\tau$ result,
Eq. \eqref{eq:abs_cs-result}, is consistent with Standard Model assuming lepton universality.
As discussed in Section \ref{sec:numuCC}, the flux of
neutrinos in the DONuT beam is approximately equal to
the flux of antineutrinos, which has been assumed for the results given 
above. The actual value of the ratio of $\bar \nu _\mu$ and
$\nu _\mu$ fluxes in the DONuT beam was measured to be $1.05 \pm 0.13$.
This $\nu$-$\bar \nu$ imbalance taken at face value would result in
a negligible correction to the  relative cross section if one assumes
that it applies to all flavors equally. The absolute cross section
would be reduced by about 2.5\%.

\section{Conclusions}
\label{sect:conclude}
  
The identification of a set of likely $\nu_{\tau}$ interactions with small background has enabled a first direct measurement of the $\nu_{\tau}$ charged-current cross section. The values obtained are consistent with the Standard Model expectation of unity for the relative cross sections. Since the uncertainty from hadronic charm production and decay is larger than the statistical error, these results can be improved with
better data from charm production experiments.

\section{Acknowledgments} 

We would like to thank the support staffs at Fermilab and the collaborating
institutions. We acknowldge the support of the U.S. Department of Energy,
the Japan Society for the Promotion of Science, the Japan-US Cooperative
Research Program for High Energy Physics, the Ministry of Education,
Science and Culture of Japan, the General Secretariat of Research and
Technology of Greece, the Korean Research Foundation, and the DOE/OJI Program.

\clearpage






\begin{table}[hbt]
\centering
\caption{Summary of the prior probabilities for the multivariate analysis.}
\begin{tabular}{ccccc}
Material & Number of && Prior probabilities ${\mathcal W}$ \\
& decay prongs & Tau decay & Charm decay & Hadron int. \\
\hline
Emulsion & 1 & $2.7\times10^{-3}$ & $1.9\times10^{-3}$ & $4.1\times10^{-5}$\\
Emulsion & 3 & $2.7\times10^{-3}$ & $1.9\times10^{-3}$ & $2.0\times10^{-4}$\\
Plastic & 1 & $1.6\times10^{-2}$ & $1.2\times10^{-3}$ & $7.5\times10^{-6}$\\
Plastic & 3 & $2.7\times10^{-3}$ & $1.9\times10^{-3}$ & $6.7\times10^{-5}$\\
Steel   & 1 & $1.6\times10^{-2}$ & $1.2\times10^{-3}$ & $5.1\times10^{-4}$\\
Steel   & 3 & $1.6\times10^{-2}$ & $1.2\times10^{-3}$ & $5.6\times10^{-3}$\\
\end{tabular}
\label{tbl:priors}
\end{table}

\begin{table}[hbt]
\centering
\caption{
Efficiencies for identifying the secondary vertex in $\nu_\tau$ interactions,
in charm-producing $\nu_e$ and $\nu_\mu$ interactions, and in $\nu$ NC events with secondary hadronic interactions. 
(Kink-daughter type is given in parentheses.) }
\begin{tabular}{lccccc}
Decay Topology  &
$\nu_\tau \rightarrow \tau^-$ & $\bar \nu_\tau \rightarrow \tau^+$ &
$\nu \rightarrow charm$ & $\bar\nu \rightarrow charm$ &  Hadron interactions\\
\hline
 1-prong (Hadron)   & 0.39 & 0.39 & 0.26 & 0.32 & 0.72\\
 1-prong (Electron) & 0.49 & 0.51 & 0.35 & 0.36 \\
 1-prong (Muon)     & 0.50 & 0.54 & 0.34 & 0.33 \\
 3-prong decay   & 0.58 & 0.62 & 0.45 & 0.56 & 0.84\\
 All             & 0.46 & 0.47 & 0.34 & 0.40 & 0.76\\
\end{tabular}
\label{tbl:taueff}
\end{table}

\begin{table}[hbt]
\centering
\caption{Expected composition of the beamdump neutrino beam. 
The distinction of $\nu_\mu$ from prompt (charm decay) and non-prompt
 ($\pi$ and K decay) sources can be made only for Monte Carlo.
 The $\rm NC_{\rm{{\it eff}}}$ category includes all events not classified as charged-current.}
\begin{tabular}{lccccc}
 & $\nu_e$ CC & $\nu_\mu$ CC & $\nu_\mu$ CC & $\nu_\tau$ CC & ${\rm NC_{\rm{{\it eff}}}}$ \\
 & & prompt & non-prompt \\
\hline
MC fraction   & 0.181  & 0.199 & 0.159 & 0.018 & 0.442 \\
MC fraction $\times$ 578  & 105 & 115 & 92 & 10 & 256 \\
Data & 120 & \multicolumn{2}{c}{225} & 9 & 224 \\ 
Difference & $15\pm15$ & \multicolumn{2}{c}{$18\pm21$} & $-1\pm4$ & $-32\pm22$ \\
\end{tabular}
\label{tbl:expect}
\end{table}

\begin{table}[hbt]
\centering
\caption{Results of a systematic study of classifying $\nu_e$ CC events as a function of electron energy. $N_e^{\rm data}$ includes both $\nu_e$ CC events and a background of NC$_{\rm{\it eff}}$ events misidentified as $\nu_e$ CC events. The last column gives the estimated true number of
CC $\nu_e$ events after subtracting background and correcting for efficiency, and should be
constant in energy if systematics are small. Events with energy less than 20 GeV were rejected
from the CC $\nu_e$ set and therefore assigned to the NC$_{\rm{\it eff}}$ set.}
\begin{tabular}{cccccc}
Energy cut & $N_e^{\rm data}$ & $N^{\rm data}({\rm NC_{\rm{{\it eff}}}})$ & $\varepsilon(\nu_e \rm{CC})$ & NC$_{\rm{\it eff}}$ bkg &$ N_e^{\rm corr}$\\
(GeV) & & & & & \\
\hline
15&144&207&0.747&0.239&154\\
18&134&217&0.693&0.194&161\\
20&120&231&0.635&0.174&166\\
25&104&247&0.573&0.160&163\\
30&91&260&0.514&0.153&165\\
\end{tabular}
\label{tbl:electron_ID}
\end{table}

\begin{table}[hbt]
\centering
\caption{List of $\nu_\tau$ events with parameters used in the  analyses and
the result of the multivariate analysis.
($\dagger$)Event 3139/22722 was a Short decay so the probability values
listed are at the tau probability minimum.}
\begin{tabular}{lcccccccccc}
Event & Daughter & $L_{dec}$ & $\alpha$ & $b_d$ & $\Delta\phi$ &
 $ \theta_p$ & $p_d$ & $P(\tau)$ & $P(c)$ & $P(int)$ \\
& & (mm) & (rad) & ($\mu$m) & (rad) & (rad) & (GeV/$c$)  \\
\hline
3024/30175  &e& 4.47 & 0.093 & 416 & 1.09&  0.030 & 5.2 & 0.53 & 0.47  & 0.00 \\
3039/01910  && 0.28 & 0.089 &  24 & 2.71 & 0.065 & 4.6 & 0.96 & 0.04  & 0.00 \\
3140/22143  &$\mu$& 4.83 & 0.012 &  60 & 1.67 &0.040 & 22.2 & 0.97 &  0.03 & 0.00 \\
3333/17665  &e& 0.66 & 0.011 &   8 & 2.84 &0.016  & 59 & 0.98 & 0.02  & 0.00 \\
3024/18706  &e& 1.71 & 0.014 &  23 & 2.96 &0.043 & 50 & 1.00 & 0.00 &  0.00 \\
3139/22722$\dagger$  && 0.44 & 0.027 &  12 & 1.71 & 0.155 &15.8  &  0.50 & 0.29 & 0.21 \\
3296/18816  && 0.80 & 0.054 & 38  &  1.74 &0.140 & 5.0 & 0.71 & 0.29  & 0.00 \\
& &   &           0.190 & 148 & & &1.3  \\
& &   &            0.130 & 112 & & &  1.9 \\
3334/19920 && 8.88 &  0.017 & 147 & 3.11 &0.041 &11.6  & 1.00 & 0.00  & 0.00 \\
& &   &            0.011 & 98 & & & 15.7 \\
& &   &             0.011  & 94 &  & & 3.2 \\
3250/01713 && 0.83 &  0.133 & 110 & 2.83 &0.028 & 1.3 & 0.87 & 0.12 &  0.01 \\
& &  &              0.192 & 161 & & & 2.4 \\
& &  &              0.442 & 355 & & &  0.5 \\
\hline
Total &&&&&&&& 7.5 & 1.26 & 0.22 \\
\end{tabular}
\label{tbl:nutau}
\end{table}


\begin{table}[hbt]
\centering
\caption{
Quantities used in the analysis 
to compute neutrino cross sections. The charm production cross section 
in a material of atomic number $A$ is assumed to be proportional
to $A^\alpha$. The differential cross section is assumed
to be given by Eq.\eqref{eq:charm_cs}.
}
\begin{tabular}{cc}
  Quantity & Value \\
\hline
 $\sigma$(pN $\rightarrow$ D$^\pm$X)  & 21 $\pm$ 2 $\mu$b \\
$\sigma$(pN $\rightarrow$ D$^0$X)     & 39 $\pm$ 3 $\mu$b \\
$\sigma$(pN $\rightarrow$ D$_s$X)     &  7.9 $\pm$ 1.2 $\mu$b \\
$\sigma$(pN $\rightarrow$ $\Lambda_c$X)     &  8 $\pm$ 5 $\mu$b \\
$\sigma_{tot}$(pW)                    &  1650 mb \\
$\alpha$ & $0.99 \pm  0.03 $ \\
$n$ & $8.0 \pm 0.8$ \\
$b$ & $0.83 \pm 0.22$ (GeV/$c$)$^{-2}$
\end{tabular}
\label{tbl:xsect-input}
\end{table}

\begin{table}[hbt]
\centering
\caption{
Monte Carlo derived factors in the cross section analysis.
}
\begin{tabular}{ccc}
 Type & $C_\ell$ & $f\left\langle {\sum {EKTt} } \right\rangle_\ell$ \\
 & $\times10^{40}$ $\rm{cm}^{-2}$ & GeV\\
\hline
$\nu_e$   & $1.64\pm0.38$ & $4.62_{-0.94}^{+1.41}$ \\
$\nu_\mu$ &$1.55\pm0.36$ & $4.33_{-0.88}^{+1.32}$\\
$\nu_\tau$  & $0.222\pm0.085$ & $2.23_{-0.52}^{+0.69}$ \\
\end{tabular}
\label{tbl:A1}
\end{table}

\begin{table}[hbt]
\centering
\caption{The values for the factors of Eq. \eqref{eq:CS5} giving the relative
 cross sections. The number of observed $\nu_\tau$ interactions is the sum
 of the probabilities  listed in Table \ref{tbl:nutau}, column 7.
The values of $C$ and $f\left\langle {\sum {EKTt} } \right\rangle$,
columns four and five, are listed in Table \ref{tbl:A1}. }
\begin{tabular}{lcccc}
$x$&${N_{\nu_x  }^{\rm{exp}}}$&
$\varepsilon \left( {\nu_x  } \right)/\varepsilon \left( {\nu_\tau }  \right)$
&$C_{x}/C_\tau$&
$f\left\langle {\sum {} } \right\rangle_x$/$f\left\langle {\sum  {} }
\right\rangle_\tau$\\
\hline
$\tau$ & 7.5 \\
$e$   &  99 & $1.36 \pm 0.08$ & $7.40\pm3.25$ & $2.07\pm0.78$\\
$\mu$ & 138 & $1.57 \pm 0.10$ & $7.01\pm2.98$ & $1.94\pm0.71$\\
\end{tabular}
\label{tbl:cs_factors}
\end{table}

\clearpage

\appendix

\section{Charm and tau production in 800-GeV proton-nucleon interactions}
\label{sec:Charm_production}

The majority of the neutrino flux at the DONuT emulsion target originated in charm decays from
interactions of 800 GeV protons in the tungsten alloy beamdump. This flux was estimated from
results of hadronic charm production in fixed-target experiments. 
The results from three experiments were used in the following way.
First, we fix the absolute rate of charm production in 800 GeV proton-nucleon
using inclusively produced D$^0$ cross sections from Ref. 
\cite{ref63}\cite{ref66}\cite{ref:HeraB}\cite{ref:charm_rev06}. 
The value of the D$^0$ cross section from \cite{ref:HeraB} was scaled from
920 GeV to 800 GeV, a factor of 0.84, using Pythia with CTEQ6L structure functions before 
averaging \cite{ref:charm_rev06}.

 We then
make the assumption that the ratio of any charm particle cross section to D$^0$ from the same
experiment is independent of energy and beam particle. The product of the weighted average of these ratios and the 800 GeV D$^0$ cross section gives our estimate for the inclusive production cross 
sections for D$^{\pm}$, D$_s$. Table \ref{tbl:charm_cs_list} lists the experimental
results used in this analysis. Table \ref{tbl:cs_ratios} gives the values for the ratios of charm species
to D$^0$ used. Note that the ratio of the $\nu_\mu$ to $\nu_e$ CC cross section ratio does
not depend on the numbers used in Table \ref{tbl:cs_ratios}. Input values used in the
cross section analysis, including charm cross sections are listed in Table \ref{tbl:xsect-input}.

The simulated charm produced in the beamdump are forced to decay semi-leptonically (or
leptonically) with the branching fractions listed in Table \ref{tbl:br_ratios}.
The charm was produced in the Monte Carlo using the simple form of Eq. \eqref{eq:charm_cs}, with
values of $n$  given in Table \ref{tbl:charm_n}. The value of $b$ was set to $0.9 \pm 0.1$.

\begin{figure}
\includegraphics[width=\columnwidth]{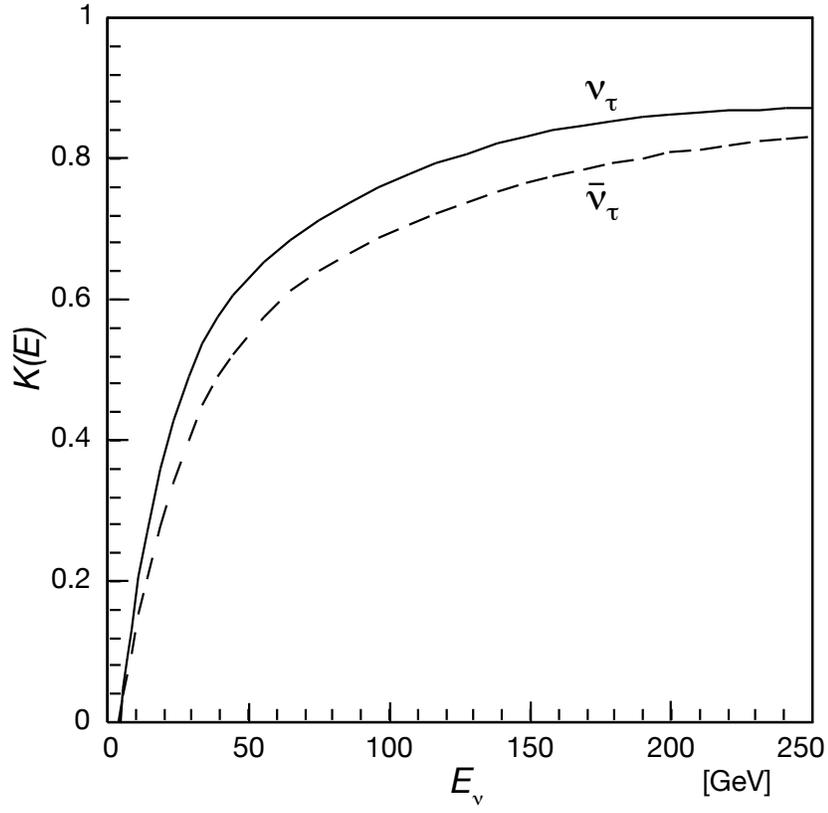}
\caption{
The tau lepton mass suppresses the $\nu_\tau$ CC cross section relative to the
$\nu_\mu$ and $\nu_e$ cross sections.}
\label{fig:kinfac}
\end{figure}

The simulation of charm production, described above, is appropriate for 800 GeV pN
interactions. Charm particles were also produced in hadronic cascade showers in the
beamdump, which we call secondary charm production. This secondary production was
modeled by the Monte Carlo in a manner similar to  non-prompt  neutrino generation.
Instead of simulating decays of $\pi$s and $K$s after each GEANT step, a charmed meson was generated and
weighted according to production cross sections via an energy-dependent function similar to
the $K(E)$ function shown in Fig. \ref{fig:kinfac}. The number of neutrino interactions from
secondary charm decays relative to total was estimated to be $0.075\pm0.033$. This 
value was applied as a correction to the absolute cross section and was assumed to
be independent of flavor.

\begin{table}[hbt]
\centering
\caption{The charm cross section results used in the cross section ratios given
in Table \ref{tbl:cs_ratios}, below. The $\rm{D}^0$ cross section was obtained from
the first three results, $pN$ reactions at high energy. The ratio of $\rm{D}^{\pm}$ to
$\rm{D}^0$ was obtained from the results above the line (all $pN$ reactions). The
ratio of $\rm{D}_s$ to $\rm{D}^0$ was obtained the results below the line. The resulting
cross sections are listed in Table \ref{tbl:xsect-input}.
}
\begin{tabular}{ccccc}
Ref. & Beam type/& $\sigma(\rm{D}^\pm)$ & $\sigma(\rm{D}^0)$ & $\sigma(\rm{D}_s)$ \\
& Energy (GeV) & $\mu$b/nucl & $\mu$b/nucl & $\mu$b/nucl \\
\hline
\cite{ref63} & p/800       & $37  \pm 9 \pm 12$   & $43 \pm 3 \pm14$      &         NA              \\
\cite{ref66} & p/800      &  $26  \pm 4 \pm 7$       & $22   \pm 8 \pm 6$      &        NA               \\
\cite{ref:HeraB} & p/920 & $ 29.9 \pm 4.5 \pm 5.7$ & $ 56.3 \pm 8.5 \pm 9.5$  &    NA \\
\cite{ref68} & p/250      &  $3.3 \pm 0.4 \pm 0.4$   & $6.0   \pm 1.4 \pm 0.5$     &    $1.5\pm1.5$     \\
\hline
\cite{ref65} & $\pi$/230 & $3.2 \pm 0.2 \pm 0.7$ & $6.6 \pm 0.3 \pm 1.0$   & $2.7 \pm 0.2$  \\
\cite{ref68} & $\pi$/250&  $3.6 \pm 0.2 \pm 0.3$   & $8.7   \pm 0.7 \pm 0.6$     &    $2.0\pm0.5$     \\
\cite{ref68} & K/250     &  $3.0 \pm 0.4$   & $7.2   \pm 1.1$     &    $3.0\pm0.9$     \\
\cite{ref68} & p/250      &  $3.2 \pm 0.5$   & $5.4   \pm 1.4$     &    $1.5\pm1.5$     \\
\cite{ref70} & $\pi$/350&  $3.2 \pm 0.1 \pm 0.3$   & $7.8   \pm 0.14 \pm 0.5$     &    $1.3\pm0.4$     \\

\end{tabular}
\label{tbl:charm_cs_list}
\end{table}

\begin{table}[hbt]
\centering
\caption{The weighted average ratio of $D^\pm$ and D$_s$ cross sections  to $D^0$ for results
listed in Table \ref{tbl:charm_cs_list}.}
\begin{tabular}{cc}
            Avg.$\frac{{\sigma \left( {D^ \pm  } \right)}}{{\sigma \left( {D^0 } \right)}}$
        & Avg.$\frac{{\sigma \left( {D_s  } \right)}}{{\sigma \left( {D^0 } \right)}}$   \\
\hline
 $0.51 \pm 0.06$  &  $0.203 \pm 0.031$ \\
\end{tabular}
\label{tbl:cs_ratios}
\end{table}

\begin{table}[hbt]
\centering
\caption{The production parameter $n$ of Eq. \eqref{eq:charm_cs} used to generate
charm particles in the Monte Carlo. The error on the values gives the range of $n$ 
used in the estimating the systematic uncertainty.}
\begin{tabular}{cc}
            Charm particle &  $n$ \\
\hline
 $\rm{D}^0$ &  $6.0 \pm 0.6$ \\
 $\overline{\rm{D}}^0$ &  $7.0 \pm 0.6$ \\
 $\rm{D}^+$ and $\rm{D}^-$ &  $5.0 \pm 0.6$ \\
 $\rm{D}_s^+$ and $\rm{D}_s^-$ &  $8.0 \pm 0.8$ \\
 $\rm{\Lambda}_c^+$ &  $2.5 \pm 0.5$ \\
 $\rm{\Lambda}_c^-$ &  $8.0 \pm 2.0$ \\
\end{tabular}
\label{tbl:charm_n}
\end{table}


\begin{table}[hbt]
\centering
\caption{Leptonic braching fractions of charm and tau used in the analysis}
\begin{tabular}{lc}
\hline
 $BR(D_s\rightarrow \nue X)$  &  $0.08\pm0.055$  \\
 $BR(D_s\rightarrow \nutau X)$  &  $0.064\pm0.015$  \\
 $BR(D_s\rightarrow \numu X)$  &  $0.08\pm0.055$  \\
 \hline
 $BR(D^{\pm}\rightarrow \nue X)$  &  $0.172\pm0.019$  \\
 $BR(D^{\pm}\rightarrow \nutau X)$  &  $7\times10^{-4}$ \\
 $BR(D^{\pm}\rightarrow \numu X)$  &  $0.16\pm0.03$ \\
 \hline
 $BR(D^{0}\rightarrow \nue X)$  &  $0.069\pm0.003$  \\
 $BR(D^{0}\rightarrow \numu X)$  &  $0.066\pm0.008$  \\
 \hline
 $BR(\Lambda_{c}\rightarrow \nue X)$  &  $0.021\pm0.007$ \\
 $BR(\Lambda_{c}\rightarrow \numu X)$  &  $0.020\pm0.006$  \\
 \hline
 $BR(\tau \rightarrow \nue X)$  &  $0.1784\pm0.0006$  \\
 $BR(\tau \rightarrow \numu X)$  &  $0.1736\pm0.0006$  \\
\end{tabular}
 \label{tbl:br_ratios}
 \end{table}

\clearpage

\end{document}